\documentclass[aps,pra,10pt,twocolumn,floatfix,superscriptaddress, longbibliography,groupedaddress]{revtex4-2}
\usepackage[T1]{fontenc}
\usepackage[colorlinks=true,urlcolor=blue,linkcolor=blue,citecolor=blue]{hyperref}
\usepackage{graphicx,overpic,orcidlink,amssymb,bm,physics}
\usepackage{subcaption}
\usepackage{amsmath}
\usepackage{amsthm}
\usepackage{mathtools}
\usepackage{booktabs}
\usepackage{siunitx}
\newtheorem{definition}{Definition}

\DeclareMathOperator{\sgn}{sgn}

\begin{document}

\title{Efficiently Simulable Pauli Correlation Encoding}

\author{Daniele {Lizzio Bosco}, Gabriel Matos, Chen-Yu Liu, Frederic Rapp, \\Fabian Finger, Enrico Rinaldi, Konstantinos Meichanetzidis}

\affiliation{Quantinuum, Partnership House, London, UK}

\begin{abstract}
Pauli Correlation Encoding (PCE) is a heuristic framework for binary optimisation that encodes classical variables into many-body Pauli observables.
While PCE requires fewer qubits than other approaches, it relies on estimating a large number of Pauli expectation values whose signs determine the variables' values, which can incur substantial measurement overhead.
Here, we introduce \emph{efficiently simulable PCE}, a class of dequantised PCE realisations where all expectation values needed can be computed efficiently classically.
We instantiate this idea using free-fermionic evolutions, realised by matchgate circuits, and Instantaneous Quantum Polynomial (IQP) circuits.
On MaxCut, Maximum Independent Set, Multi-Dimensional Knapsack, and Max3SAT benchmarks, these methods produce high-quality solutions across problem sizes ranging from tens to thousands of variables.
Our results show that PCE is naturally understood as a correlation-based optimisation framework with both quantum and classically simulable realisations.
This yields a dequantised baseline for evaluating future quantum PCE implementations.
\end{abstract}

\maketitle

\section{Introduction}

Quantum algorithms for combinatorial optimisation are often motivated by the ability of parameterised quantum circuits to generate complex correlation structures that may be difficult to reproduce classically~\cite{farhi2014quantumapproximateoptimizationalgorithm, PhysRevA.95.032323, PhysRevApplied.19.024027}. Such correlations have been explored in a variety of variational and hybrid quantum-classical approaches~\cite{liu2022hybridgatebasedannealingquantum, Tan_2021, Tilly_2022}, where the optimisation process searches over a family of parameterised quantum states. Understanding whether these correlations can provide practical computational advantages remains one of the central challenges in quantum optimisation.

Pauli Correlation Encoding (PCE) is a recently proposed quantum heuristic framework designed to address large-scale combinatorial optimisation problems~\cite{sciorilli2024largescalequantumoptimizationsolvers}. Rather than associating each optimisation variable directly with a qubit, PCE encodes binary variables into the expectation values of Pauli observables. A parameterised quantum circuit generates a quantum state, and the signs of the resulting expectation values determine the values of the optimisation variables. This encoding allows a reduction in the number of qubits required to represent a problem instance while retaining a variational optimisation loop.

The role of the quantum circuit within PCE is not to generate samples from a computationally hard distribution, but to produce correlations from which candidate solutions can be decoded. Although the homogeneous encoding of Ref.~\cite{sciorilli2024largescalequantumoptimizationsolvers} requires only three measurement settings, repeatedly estimating these correlations remains a bottleneck. In particular, resolving the sign of a correlation of magnitude $\gamma$ with fixed confidence requires $\mathcal{O}(\gamma^{-2})$ shots; resolving all $m$ signs with failure probability at most $\delta$ adds a factor $\log(m/\delta)$ under standard concentration bounds~\cite{PhysRevA.92.042303, McClean_2016, hoeffding1963probability, majsak2025jointmeasurement}. Correlations close to zero can therefore make training prohibitively expensive on quantum hardware~\cite{alonso2026benchmarkpaulicorrelationencoding}, even when the qubit count and number of measurement settings remain small.

Interestingly, the reliance of PCE on expectation values rather than output samples provides considerable flexibility in the choice of the underlying circuit ansatz. Since candidate solutions are decoded entirely from observable expectation values, one may consider restricted circuit families for which the relevant correlations can be computed efficiently on classical hardware. Examples include matchgate circuits~\cite{terhal_2002, josza_2008, surace_2022, matos2023}, which realise free-fermionic evolutions, and Instantaneous Quantum Polynomial (IQP) circuits~\cite{Bremner_2010, leontica_2024}, both of which admit polynomial-time algorithms for evaluating specific classes of expectation values.

This observation raises a broader question regarding the practical utility of quantum implementations of PCE. If a restricted circuit family can generate useful optimisation correlations while allowing efficient classical estimation of the encoded observables, then it naturally provides a dequantised version of the same optimisation pipeline. Consequently, a rigorous assessment of quantum utility or advantage in PCE should not only compare against conventional classical optimisation algorithms, but also against classically simulable realisations of the PCE framework itself. This perspective has proven valuable in other areas of quantum computing. Prior work has successfully employed matchgate and IQP circuit families to characterise the performance of variational quantum algorithms, investigate the capabilities of QAOA-like ansatz, and develop scalable approaches to quantum machine learning~\cite{matos2023,gince2024,leontica_2024,recioarmengol2026trainclassicaldeployquantum,liu2026generativequantumutilitycorrelationcomplexity}. These results suggest that classically simulable circuit families can serve both as practically useful optimisation models and as important baselines for evaluating genuinely quantum approaches.

In this work, we investigate this idea within the PCE framework. We introduce a class of \emph{efficiently simulable PCE} realisations, in which the variational ansatz and encoded observables are chosen such that the correlations required for decoding solutions can be computed in polynomial time on classical hardware. We study two representative instances of this framework: one whose ansatz is a matchgate circuit realising free-fermionic evolution, which we refer to as \emph{free-fermionic PCE} (FF-PCE), and one based on IQP circuits (IQP-PCE). We use ``free fermions'' as the general label for this construction and ``matchgate circuit'' for the specific ansatz that implements it, adopting the shorthand FF-PCE throughout.
By operating entirely classically, these variants can address optimisation problems containing up to several thousand variables across benchmark tasks including MaxCut, Maximum Independent Set, Multi-Dimensional Knapsack Problem, and Max3SAT, while achieving competitive solution quality.

Taken together, these results position efficiently simulable PCE as a useful point of comparison for future quantum implementations.
Because the dequantised variants retain the same form, optimisation loop, and sign-based decoding rule, they help separate the contribution of the PCE framework itself from the additional expressive power and measurement cost of a more general quantum ansatz~\cite{pennylaneBenchmarkingQuantum}. In this sense, they provide a controlled setting for assessing when genuinely quantum circuit families offer an advantage within Pauli Correlation Encoding.

This work is organised as follows. In Sec.~\ref{sec:PCE}, we provide a general introduction to Pauli Correlation Encoding. In Sec.~\ref{sec:efficient-pce}, we present our efficiently simulable PCEs, with a focus on Free Fermions (\ref{subsec:ff-pce}) and IQP (\ref{subsec:iqp-pce}). Numerical results are presented in Sec.~\ref{sec:exp-results} and discussed in Sec.~\ref{sec:discussion}, where we also draw our conclusions and outlook.

\section{Pauli Correlation Encoding}
\label{sec:PCE}

Pauli Correlation Encoding (PCE) is a quantum-classical framework for binary combinatorial optimisation in which classical variables are represented by quantum correlations rather than by individual qubits~\cite{sciorilli2024largescalequantumoptimizationsolvers}. This distinction is central to the scalability motivation of PCE. In a conventional qubit-variable encoding, representing $m$ binary variables requires $\mathcal{O}(m)$ qubits. By contrast, PCE assigns variables to expectation values of Pauli observables acting on a smaller register. If observables of weight at most $k$ are used, then $\mathcal{O}(n^k)$ variables can be encoded using $n$ qubits, giving $n=\mathcal{O}(m^{1/k})$ for fixed $k$, and providing a polynomial compression of factor $k$.

\begin{figure*}[t]
    \centering
    \includegraphics[
        page=1,
        width=.95\linewidth,
    ]{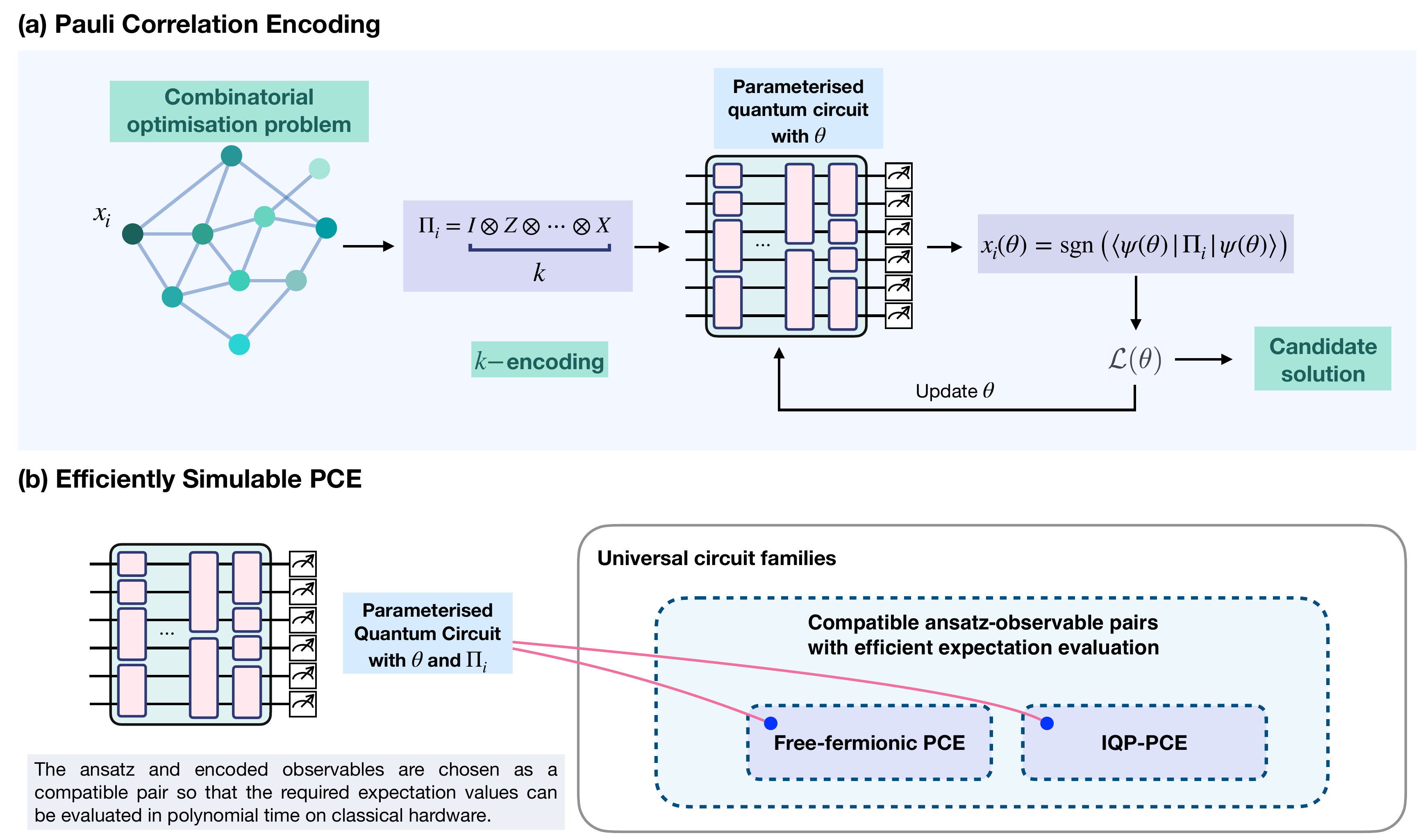}
    \caption{
    Schematic comparison between standard PCE and efficiently simulable PCE.
    (a) In standard PCE, binary variables are encoded into Pauli observables, whose expectation values are evaluated using a parameterised quantum circuit and decoded into a candidate solution.
    (b) In efficiently simulable PCE, the ansatz family and encoded observables are chosen as a compatible pair so that all expectation values required for training and decoding can be evaluated in polynomial time on classical hardware. Free-fermionic PCE and IQP-PCE are two examples of such classically simulable realisations.
    }
    \label{fig:scheme-pce}
\end{figure*}

Consider an unconstrained binary optimisation problem
\begin{equation}
\min_{x\in\{-1,1\}^m} f(x),
\end{equation}
where $x=\{x_i\}_{i<m}$.
Note that the form of binary variables can always be rewritten to  different representations such as $y_i \in \{0,1\}$ via the standard mapping
\begin{equation}
y_i := (x_i + 1) / 2, y_i\in\{0,1\}.
\end{equation}
A PCE instance is specified by a collection of $m$ traceless Pauli strings
\begin{equation}
\Pi=\{\Pi_i\}_{i<m}, \qquad
\Pi_i\in\{I,X,Y,Z\}^{\otimes n}\setminus \{I^{\otimes n}\}.
\end{equation}
Given a parameterised quantum state with ansatz $\mathcal{U}$, parameters $\theta$, and initial state $|\psi_0\rangle$:
\begin{equation}
|\psi(\theta)\rangle=\mathcal{U}(\theta)|\psi_0\rangle,
\end{equation}
the encoded variables are decoded from the signs of the corresponding expectation values:
\begin{equation}
\label{eq:PCE_assignment}
x_i(\theta) \coloneqq \sgn\left(\langle \Pi_i\rangle_\theta\right),
\qquad
\langle \Pi_i\rangle_\theta
\coloneqq
\langle \psi(\theta)|\Pi_i|\psi(\theta)\rangle .
\end{equation}
If every observable in $\Pi$ has Pauli weight at most $k$, we refer to $\Pi$ as a $k$-encoding.
Once the encoding $\Pi$ and ansatz $\mathcal{U}(\theta)$ are fixed, the original discrete optimisation problem is transformed into a continuous optimisation problem over the parameters $\theta$. Since the sign function is non-differentiable, PCE typically replaces it during training with a smooth relaxation, for example
\begin{equation}
\mathcal{L}(\theta)
=
f\left(\tanh(\alpha\langle \bm{\Pi}\rangle_\theta)\right)
+
\beta \mathcal{L}^{(\mathrm{reg})},
\end{equation}
where $\langle\bm{\Pi}\rangle_\theta=(\langle\Pi_i\rangle_\theta)_{i<m}$ denotes the vector of encoded expectation values and $\tanh$ is applied component-wise, $\alpha>0$ controls the sharpness of the relaxation and $\mathcal{L}^{(\mathrm{reg})}$ denotes a problem-dependent regularisation or penalty term. After training, a discrete candidate solution is obtained by applying the sign rule in Eq.~\eqref{eq:PCE_assignment}. As in many heuristic optimisation methods, this candidate solution may subsequently be refined by a classical local-search procedure, such as greedy single-bit flips.
This decomposition highlights the modular structure of PCE. As depicted in Fig.~\ref{fig:scheme-pce}, the PCE pipeline consists of three ingredients:
\begin{enumerate}
\item A Pauli encoding $\Pi$ assigning optimisation variables to observables.
\item A parameterised ansatz $\mathcal{U}(\theta)$ generating the correlations used for decoding.
\item A classical optimiser that updates $\theta$ using a loss function derived from the target objective.
\end{enumerate}
The first and third components are entirely classical. In a standard quantum implementation, the quantum device is used only to estimate the expectation values $\langle \Pi_i\rangle_\theta$ generated by the ansatz.
This observation is important for resource estimation. Although PCE reduces the number of qubits required to represent a large problem instance, it does not remove the cost of extracting the encoded correlations. Estimating a Pauli expectation value to additive precision $\epsilon$ requires $\mathcal{O}(\epsilon^{-2})$ measurements~\cite{PhysRevA.92.042303, McClean_2016}. However, a single batch of shots can be reused for all observables measured in the same basis. Suppose that the $m$ encoded observables can be partitioned into $G$ jointly measurable settings. Standard concentration bounds~\cite{hoeffding1963probability, majsak2025jointmeasurement} imply that all $m$ expectation values can be estimated to additive error $\epsilon$, with total failure probability at most $\delta$, using a number of shots equal to
\begin{equation}
N_{\mathrm{shots}}^{(\mathrm{loss})}
=
\mathcal{O}\!\left(
\frac{G}{\epsilon^2}\log\frac{m}{\delta}
\right)
\label{eq:grouped-shot-cost}
\end{equation}
If gradients are estimated using a parameter-shift rule~\cite{li_2017, mitarai_2018}, each trainable parameter requires $2$ shifted circuit evaluations. For an ansatz with $N_p$ trainable parameters and $N_t$ optimisation steps, the total shot cost scales as
\begin{equation}
N_{\mathrm{shots}}^{(\mathrm{train})}
=
\mathcal{O}\!\left(
N_t(c+2N_p)\frac{G}{\epsilon^2}\log\frac{m}{\delta}
\right),
\end{equation}
where $c$ denotes the number of additional unshifted loss evaluations performed per optimisation step. In many practical training routines, one may take $c=1$, for example when the current loss is evaluated for logging, early stopping, line search, or selecting the best parameters, while $c=0$ corresponds to estimating only the shifted losses required for the gradient.

To obtain a shot estimate for illustrative purposes, let $\mathcal{E}$ denote the collection of all parameter settings evaluated during optimisation, including the unshifted and parameter-shifted circuits. Define the minimum correlation magnitude over every encoded sign and every evaluation,
$
\gamma_{\min}=\min_{\theta\in\mathcal{E}}\min_{i<m}\left|\langle\Pi_i\rangle_\theta\right|>0.
$
Taking $m=10^3$ encoded variables, $N_p=10^2$ trainable parameters, $N_t=10^3$ optimisation steps, $c=1$, $G=3$, $\delta=10^{-2}$, and $\gamma_{\min}=10^{-2}$ gives $N_{\mathrm{shots}}^{(\mathrm{train})}\approx5.9\times10^{11}$. Even at an effective shot time of $t_{\mathrm{shot}}=40\,\mu\mathrm{s}$, this corresponds to approximately $273$ days of continuous serial circuit execution;  estimates for other margins are given in Appendix~\ref{app:wall-clock}.
If one or more correlations approach zero, this cost diverges even for fixed $G$. This correlation-margin dependence, compounded by repeated circuit evaluations during optimisation, is the principal measurement challenge in PCE. Adaptive shot allocation can reduce expenditure on correlations whose signs are already clear, but it cannot remove the inverse-square cost of resolving the small correlations that matter to the decoded solution.

The central question addressed in this work is therefore whether the correlation-generation stage of PCE must necessarily be implemented by a quantum circuit whose relevant expectation values are hard to compute classically. In the following sections, we show that by choosing compatible pairs of ansatz families and encoded observables, one obtains PCE realisations in which all correlations required for training and decoding can be evaluated efficiently on classical hardware. Classical simulability by itself implies no ordering of solution quality: a restricted, efficiently simulable ansatz may represent fewer correlation patterns than a more general circuit, but the quality obtained in practice also depends on whether the correlations relevant to the problem can be represented and reliably optimised within the available resource budget. Efficiently simulable PCE can therefore perform better, worse, or comparably, depending on the ansatz, problem, and optimisation protocol. We examine this question empirically in Sec.~\ref{sec:exp-results}, including a controlled comparison with a linear-entangling PCE baseline in Fig.~\ref{fig:pce_cobyla_comparison}.

\section{Pauli Correlation Encoding with Efficiently Simulable Circuits}
\label{sec:efficient-pce}

The discussion above highlights that the quantum subroutine in a standard PCE pipeline is used only to evaluate the encoded correlations \(\{\langle \Pi_i\rangle_\theta\}_{i<m}\). This observation motivates a natural modification of the framework. Instead of requiring the ansatz \(\mathcal{U}(\theta)\) to be a generic, potentially hard-to-simulate quantum circuit, one may choose the ansatz and the encoding jointly so that all expectation values required for training and decoding can be evaluated efficiently on classical hardware.

This compatibility between the circuit family and the encoded observables is essential. Some classically simulable circuit families allow efficient evaluation of a broad class of Pauli observables. For example, Clifford circuits can be tracked efficiently using the stabilizer formalism, and therefore Pauli expectation values can be computed in polynomial time~\cite{gottesman1998heisenbergrepresentationquantumcomputers, PhysRevA.70.052328}. Other circuit families, such as IQP circuits, lead to efficiently simulable observables composed of Pauli $Z$ operators. In such cases, the PCE encoding must be chosen to lie within the efficiently accessible set of correlations. Thus, efficient simulability is not solely a property of the ansatz, but of the pair consisting of the ansatz family and the chosen Pauli encoding.

We exploit this observation to construct classically simulable realisations of PCE. These realisations preserve the high-level structure of the original framework: variables are still encoded into Pauli expectation values, the ansatz is still trained by optimising a relaxed objective, and candidate solutions are still decoded from the signs of the resulting correlations. The \textit{difference} is that the expectation values are computed directly by a polynomial-time classical algorithm rather than estimated from quantum measurements. A schematic comparison between the standard PCE pipeline and the efficiently simulable variant is shown in Fig.~\ref{fig:scheme-pce}.

This perspective serves two purposes. First, it yields quantum-inspired classical optimisation heuristics that inherit the correlation-based encoding structure of PCE while avoiding the measurement overhead of quantum hardware. Second, it provides dequantised baselines for assessing future quantum implementations of PCE.
A quantum ansatz can offer practical utility only if its additional expressivity produces optimisation improvements that cannot already be obtained from efficiently simulable correlation models and that are large enough to compensate for the cost of quantum expectation-value estimation.

\subsection{Efficiently Simulable Pauli Correlation Encoding}

To make this compatibility condition precise, we define an efficiently simulable PCE as a choice of encoding and ansatz for which all correlations required by the PCE pipeline can be evaluated efficiently on classical hardware.
\begin{definition}[Efficiently simulable PCE]
\label{def:efficiently-simulable-pce}
Let $\Pi=\{\Pi_i\}_{i<m}$ be a collection of Pauli observables acting on $n$ qubits, with $m=|\Pi|\in\mathrm{poly}(n)$, and let $\mathcal{U}(\theta)$ be a parameterised ansatz preparing the state
\begin{equation}
    |\psi(\theta)\rangle = \mathcal{U}(\theta)|\psi_0\rangle .
\end{equation}
We say that the pair $(\Pi,\mathcal{U})$ defines an \emph{efficiently simulable Pauli Correlation Encoding} if there exists a classical algorithm that, for every $\Pi_i\in\Pi$ and every inverse-polynomial precision $\epsilon=1/\mathrm{poly}(n)$, outputs an estimate $\widetilde{\mu}_i(\theta)$ satisfying
\begin{equation}
    \left|
    \widetilde{\mu}_i(\theta)
    -
    \langle \Pi_i\rangle_\theta
    \right|
    \leq \epsilon,
\end{equation}
with runtime polynomial in $n$ and $1/\epsilon$, where
\begin{equation}
    \langle \Pi_i\rangle_\theta
    \coloneqq
    \langle \psi(\theta)|\Pi_i|\psi(\theta)\rangle .
\end{equation}
\end{definition}

The condition \(m=|\Pi|\in\mathrm{poly}(n)\) ensures that the number of encoded variables remains polynomial in the number of qubits. This rules out trivial cases in which a circuit family allows efficient access to only too few observables to define a meaningful compressed encoding. Conversely, the definition does not require that all Pauli observables be efficiently computable. It is sufficient that the particular observables used for the PCE encoding are efficiently accessible for the chosen ansatz family.

In the remainder of this section, we instantiate this definition using two representative classes of efficiently simulable PCE:
\begin{itemize}
    \item \textbf{Free-Fermionic PCE.} We consider ansatz circuits representing free-fermionic evolutions. Using the matchgate representation, the relevant correlations are obtained from the corresponding Majorana covariance matrix. By encoding variables into its independent two-point correlations, this construction yields a quadratic compression of variables into qubits.

    \item \textbf{IQP-PCE.} We consider Instantaneous Quantum Polynomial (IQP) circuits composed of mutually commuting rotations of the form $R_{\mathbf{X}}(\theta)$ for $\mathbf{X}\in\{I,X\}^{\otimes n}$. For suitable choices of encoded observables, this structure allows efficient classical evaluation of $Z$-type correlations $\langle \mathbf{Z}\rangle$ with $\mathbf{Z}\in\{I,Z\}^{\otimes n}$. This enables efficient training of PCE models with structured $k$-encodings.
\end{itemize}
These two examples illustrate different ways in which restricted, classically tractable correlation models can be embedded into the PCE framework. Note that these are not exhaustive, and other families of classically simulable ansatze can be used.

\subsection{Free-fermionic PCE}
\label{subsec:ff-pce}

Our first efficiently simulable PCE construction is based on free-fermionic dynamics~\cite{terhal_2002, josza_2008, surace_2022, matos2023}, which we denote by FF-PCE. We now briefly describe the construction used to represent and compute these dynamics efficiently.
First, we say that a set of operators $\{\gamma_j\}_{j=1,\ldots,2n}$ is called a set of Majorana operators if it satisfies the anticommutation relations
\begin{equation}
    \{\gamma_p,\gamma_q\}=2\delta_{pq}I.
\end{equation}
These operators can be mapped to Pauli observables via a fermion-to-qubit mapping. The most common of these mappings is the Jordan--Wigner transformation~\cite{JordanWigner1928, lieb_1961, chapman_2020}, for instance
\begin{eqnarray}
\label{eq:jordan_wigner}
    &&\gamma_{2j-1}
    :=
    Z_1 Z_2\cdots Z_{j-1}X_j, \nonumber \\
    &&\gamma_{2j}
    :=
    Z_1 Z_2\cdots Z_{j-1}Y_j, \nonumber \qquad j=1,\ldots,n.
\end{eqnarray}

A free-fermion Hamiltonian is a Hamiltonian that can be written as
\begin{equation}
    H_{\mathrm{FF}}
    =
    \frac{\mathrm{i}}{4}
    \sum_{p,q=1}^{2n}
    A_{pq}\gamma_p\gamma_q,
    \qquad
    A^\mathsf{T}=-A,
    \qquad
    A\in\mathbb{R}^{2n\times 2n}.
\end{equation}
The unitary it generates $U_{\mathrm{FF}} = e^{-\mathrm{i}H_{\mathrm{FF}}}$ is called a fermionic Gaussian unitary, and it acts linearly on Majorana operators:
\begin{equation}
\label{eq:orthogonal_action}
    U_{\mathrm{FF}}^\dagger \gamma_p U_{\mathrm{FF}}
    =
    \sum_{q=1}^{2n}
    O_{pq}\gamma_q,
    \qquad
    O=e^A \in SO(2n).
\end{equation}
Note that products of pairs of Majoranas under the Jordan-Wigner transformation~\eqref{eq:jordan_wigner} are Pauli matrices, which we denote by quadratic Pauli observables
\begin{align}
Z_j
&=
-\mathrm{i}\gamma_{2j-1}\gamma_{2j}, \nonumber
\\
Y_j Z_{j+1}\cdots Z_{k-1}X_k
&=
\mathrm{i}\gamma_{2j-1}\gamma_{2k-1}, \nonumber
\\
Y_j Z_{j+1}\cdots Z_{k-1}Y_k
&=
\mathrm{i}\gamma_{2j-1}\gamma_{2k}, \nonumber
\\
X_j Z_{j+1}\cdots Z_{k-1}X_k
&=
-\mathrm{i}\gamma_{2j}\gamma_{2k-1}, \nonumber
\\
X_j Z_{j+1}\cdots Z_{k-1}Y_k
&=
-\mathrm{i}\gamma_{2j}\gamma_{2k},
\qquad j<k. \label{eq:quadratic_paulis}
\end{align}
A matchgate~\cite{valiant_2002, josza_2008} is any unitary generated by a linear combination of nearest neighbour quadratic Pauli observables. For ~\eqref{eq:quadratic_paulis}, these are $Z_j,  X_j X_{j+1},  X_j Y_{j+1},  Y_j X_{j+1},  Y_j Y_{j+1}$.

A fermionic Gaussian state $\rho$ is fully determined, up to a phase, by its covariance matrix
\begin{equation}
    M_{pq}
    :=
    -\frac{\mathrm{i}}{2}
    \left\langle
        [\gamma_p,\gamma_q]
    \right\rangle_\rho,
    \qquad
    M^\mathsf{T}=-M.
\end{equation}
Note that the covariance matrix entries correspond to the expectation values of the Pauli observables in~\eqref{eq:quadratic_paulis}, up to constant factors. If $M_0$ is the covariance matrix of an input state $\ket{\psi_0}$, then the covariance matrix of $U_{\mathrm{FF}} \ket{\psi_0}$ is
\begin{equation}
    M
    =
    O M_0 O^\mathsf{T}.
\end{equation}
Thus, instead of propagating a $2^n$-dimensional state vector, the simulation propagates a real $2n\times 2n$ covariance matrix. This gives an efficient classical description of the correlations used by the model.

\paragraph{Ansatz}
Any possible fermionic Gaussian unitary can be decomposed as a sequence of the matchgates $R_Z(\theta)$ and $R_{XX}(\theta)$ with linear connectivity~\cite{DAlessandro2007, matos2023}. For this reason, we construct our ansatz as
\begin{equation}
    U_{\mathrm{FF}}(\bm{\theta})
    =
    \prod_{\ell=1}^{L}
    \left[
        U_{XX}^{(\ell)}
        U_Z^{(\ell)}
    \right],
\end{equation}
where
\begin{equation}
\begin{aligned}
    U_Z^{(\ell)}
    =
    \prod_{j=1}^{n}
    R_Z^{(j)}\!\left(\varphi_{\ell,j}\right), \quad
    U_{XX}^{(\ell)}
    =
    \prod_{j=1}^{n-1}
    R_{XX}^{(j,j+1)}\!\left(\phi_{\ell,j}\right),
    \qquad
\end{aligned}
\end{equation}
and
\begin{equation}
\begin{aligned}
    &R_Z^{(j)}(\varphi)
    =
    \exp\!\left(
        -\frac{\mathrm{i}\varphi}{2}Z_j
    \right),
    \\
    &R_{XX}^{(j,j+1)}(\phi)
    =
    \exp\!\left(
        -\frac{\mathrm{i}\phi}{2}X_jX_{j+1}
    \right).
\end{aligned}
\end{equation}
Note that this choice of ansatz is not unique, and different choices of matchgates can represent the full set of fermionic Gaussian unitaries, e.g., we could replace $U_{XX}$ by $U_{YY}$, $U_{XY}$, or $U_{YX}$~\cite{DAlessandro2007, knill2001}. The depth parameter $L$ controls the expressivity of the ansatz. Since each layer contains $2n-1$ parameters, the total number of trainable parameters is $L(2n-1)$. Note that $L = n$ layers are enough to represent any fermionic Gaussian unitary~\cite{kokcu_2022}.

\paragraph{Encoding}
The encoding used in FF-PCE uses the quadratic Pauli observables in the chosen fermion-to-qubit mapping, e.g. Eq.~\eqref{eq:quadratic_paulis}. The expectation values of these observables correspond to the strict upper-triangular entries of the Majorana covariance matrix. For $1\leq p<q\leq 2n$ we define $\Pi_{pq}:=-\mathrm{i}\,\gamma_p\gamma_q$ and assign
\begin{equation}
    z_{pq}(\bm{\theta})
    =
    \langle \Pi_{pq}\rangle_{\bm{\theta}}
    =
    M_{pq}(\bm{\theta})
    = -\mathrm{i} \langle \gamma_p\gamma_q\rangle_{\bm{\theta}}.
\end{equation}
Note that the number of quadratic Pauli observables is exactly
\begin{equation}
    \binom{2n}{2}
    =
    n(2n-1).
\end{equation}
Given an optimisation problem with $m$ variables, we choose the smallest number of qubits satisfying
\begin{equation}
    n(2n-1)\geq m,
\end{equation}i.e.
\begin{equation}
    n_{\mathrm{FF}}(m)
    =
    \left\lceil
        \frac{1+\sqrt{1+8m}}{4}
    \right\rceil.
\end{equation}
We then assign each variable $x_a$ to one covariance observable $\Pi_{p_aq_a}$, with $(p_a,q_a)\in\mathcal{I}_n$, and define
\begin{equation}
    z_a(\bm{\theta})
    =
    M_{p_aq_a}(\bm{\theta}),
    \qquad
    \hat{x}_a(\bm{\theta})
    =
    \sgn
    \left(
        z_a(\bm{\theta})
    \right).
\end{equation}
This construction gives a quadratic PCE compression:
\begin{eqnarray}
    &&m
    \leq
    n(2n-1)
    =
    \mathcal{O}(n^2),\nonumber \\
    &&\text{or equivalently}
    \quad
    n
    =
    \mathcal{O}(\sqrt{m}).
\end{eqnarray}

While it would be possible to compute the expectation value of any observable with respect to a fermionic Gaussian state in $\mathcal{O}(n^3)$ classical operations via Wick's theorem~\cite{wick_1950, surace_2022}, our construction encodes only the quadratic Pauli observables. This is because Wick's theorem implies that all higher-order correlations are determined by Pfaffians of submatrices of $M$. Therefore, these observables do not introduce independent variational degrees of freedom beyond those already contained in the covariance matrix. For this reason, we restrict FF-PCE to quadratic Pauli observables, obtaining an efficiently simulable PCE with a fixed quadratic compression.

An advantage of combining PCE with a free-fermion approach is that the cost function is not hard-coded into the ansatz, unlike in QAOA, where mixer unitaries alternate with cost unitaries generated by the problem Hamiltonian~\cite{farhi2014quantumapproximateoptimizationalgorithm}. Hard-coding this structure would make ansatz unitaries need to admit a representation as fermionic Gaussian unitaries, which is not possible for generic Pauli Hamiltonians. \citet{chapman_2020} characterised a generalized Jordan--Wigner construction by showing that such a free-fermion representation exists precisely when the Hamiltonian's frustration graph
is a line graph. More general mappings exist for claw-free frustration graphs containing a simplicial clique~\cite{elman_2021, chapman_2023}. These are special graph-theoretic conditions, so standard QAOA can not use a free-fermion ansatz with an arbitrary problem graph.

\paragraph{Classical complexity.}
The classical cost of FF-PCE is governed by the propagation of the \(2n\times 2n\) Majorana covariance matrix. A direct evaluation of
\begin{equation}
    M(\bm{\theta})
    =
    O(\bm{\theta})M_0O(\bm{\theta})^\mathsf{T}
\end{equation}
requires multiplying three $n \times n$ matrices, and therefore scales as \(\mathcal{O}(n^3)\) per loss evaluation~\cite{golub2013matrix}. In a gate-by-gate implementation, each local matchgate updates a linear-size block of the covariance matrix.
For the layered ansatz used here, with \(\mathcal{O}(n)\) layers and \(\mathcal{O}(n)\) gates per layer, this gives \(\mathcal{O}(n^3)\) classical operations. Note that reading out the encoded variables amounts to extracting the selected entries of \(M(\bm{\theta})\), which has no additional cost.

\subsection{IQP-PCE}
\label{subsec:iqp-pce}

Our second efficiently simulable PCE construction is based on Instantaneous Quantum Polynomial (IQP) circuits. IQP circuits form a family of commuting quantum circuits that are diagonal in the Pauli-$X$ basis. While sampling from sufficiently general IQP output distributions is conjectured to be classically intractable~\cite{Bremner_2010}, certain expectation values can nevertheless be evaluated efficiently on classical hardware~\cite{recioarmengol2026trainclassicaldeployquantum, liu2026generativequantumutilitycorrelationcomplexity}. We use the following consequence of the algorithm of Van den Nest \cite{nest2009simulating}, as stated in Ref.~\cite{recioarmengol2026trainclassicaldeployquantum}: given a parameterised IQP circuit $\mathcal{U}_{\mathrm{IQP}}(\bm{\theta})$, a Pauli-$Z$ word $\mathbf{Z}_{\bm{a}}$, and an additive error parameter $\epsilon$, there exists a classical polynomial-time and polynomial-space algorithm that samples a random variable $\widehat{z}_{\bm{a}}$ satisfying
\begin{equation}
    \mathbb{E}\!\left[\widehat{z}_{\bm{a}}\right]
    =
    \langle \mathbf{Z}_{\bm{a}}\rangle_{\bm{\theta}},
    \qquad
    \mathrm{Std}\!\left[\widehat{z}_{\bm{a}}\right]
    \leq
    \epsilon .
\end{equation}

This separation is useful for PCE: the optimisation pipeline depends on expectation values rather than on sampling from the full output distribution.
An $n$-qubit IQP circuit can be written as
\begin{eqnarray}
    &&\mathcal{U}_{\mathrm{IQP}}(\bm{\theta})
    =
    \exp\!\left(
        -\mathrm{i}
        \sum_{\alpha=1}^{r}
        \theta_\alpha \mathbf{X}_\alpha
    \right)
    =
    \prod_{\alpha=1}^{r}
    \exp\!\left(
        -\mathrm{i}\theta_\alpha \mathbf{X}_\alpha
    \right),\nonumber\\
    &&\mathbf{X}_\alpha\in\{I,X\}^{\otimes n}.
\end{eqnarray}
Here each $\mathbf{X}_\alpha$ is a Pauli-$X$ string supported on a subset of qubits. Since all such generators commute, the order of the gates is irrelevant, giving rise to the term \emph{instantaneous}. Equivalently, by conjugating with Hadamard gates, the same circuit can be viewed as a diagonal phase model in the computational basis:
\begin{eqnarray}
    &&\mathcal{U}_{\mathrm{IQP}}(\bm{\theta})
    =
    H^{\otimes n}
    \exp\!\left(
        -\mathrm{i}
        \sum_{\alpha=1}^{r}
        \theta_\alpha \mathbf{Z}_\alpha
    \right)
    H^{\otimes n},\nonumber\\
    &&\mathbf{Z}_\alpha\in\{I,Z\}^{\otimes n}.
\end{eqnarray}
This representation highlights the parity-structured nature of IQP circuits: each generator corresponds to a subset of qubits and therefore to a particular multi-qubit parity interaction. Although one could in principle include all subsets of qubits, this would require exponentially many parameters. A scalable variational ansatz must therefore select only polynomially many generators; common choices include low-weight or problem-structured terms~\cite{leontica_2024,iqpopt}.

For the PCE encoding, we choose observables from the $Z$-Pauli basis. For any subset $T\subseteq\{1,\ldots,n\}$, define
\begin{equation}
\label{eq:z_string}
    \mathbf{Z}_T
    =
    \prod_{j\in T} Z_j .
\end{equation}
The encoded expectation values are then
\begin{equation}
    \langle \mathbf{Z}_T\rangle_{\bm{\theta}}
    =
    \langle 0|^{\otimes n}
    \mathcal{U}_{\mathrm{IQP}}^\dagger(\bm{\theta})
    \mathbf{Z}_T
    \mathcal{U}_{\mathrm{IQP}}(\bm{\theta})
    |0\rangle^{\otimes n}.
\end{equation}
For IQP circuits, such $Z$-basis expectation values can be estimated in classical polynomial time \cite{recioarmengol2026trainclassicaldeployquantum}. Therefore, if the PCE loss and decoding rule depend only on a polynomial-size collection of observables $\{\mathbf{Z}_{T_i}\}_{i<m}$, the full training and decoding pipeline can be implemented without quantum measurements.
This gives an efficiently simulable PCE model by choosing
\begin{equation}
    \Pi_i = \mathbf{Z}_{T_i},
    \qquad
    |T_i|\leq k,
\end{equation}
and decoding variables according to
\begin{equation}
    x_i(\bm{\theta})
    =
    \sgn
    \left(
        \langle \mathbf{Z}_{T_i}\rangle_{\bm{\theta}}
    \right).
\end{equation}
The IQP construction differs from FF-PCE in an important way. In FF-PCE, the accessible observables are naturally fixed by the entries of the Majorana covariance matrix. In IQP-PCE, by contrast, the encoded observables and the ansatz generators can be selected independently: the observable weight controls the PCE encoding, while the generator structure controls the correlations introduced by the ansatz.

\paragraph{Problem-dependent ansatz and encoding}
This decoupling allows the ansatz to be adapted to the structure of the optimisation problem. Consider graph-based combinatorial problems such as MaxCut or Maximum Independent Set, defined on a graph
\begin{equation}
    G=(V,E),
    \qquad
    V=\{1,\ldots,m\},
    \qquad
    E\subseteq V\times V.
\end{equation}
For such problems, we use a $1$-encoding by assigning each classical variable to a single-qubit observable.
\begin{equation}
    x_i(\bm{\theta})
    =
    \sgn
    \left(
        \langle Z_i\rangle_{\bm{\theta}}
    \right),
    \qquad
    i\in V.
\end{equation}
This is the simplest variable assignment that introduces no algebraic constraints at the level of the observable encoding; for $n$ qubits, at most $n$ variables can be assigned in this way (see Appendix~\ref{app:iqp_maximum_independent_assignments}).This is analogous to the constraint imposed by the Pfaffian in FF-PCE. A restricted ansatz may nevertheless impose additional constraints on the reachable expectation values.  The ansatz is then chosen to reflect the graph topology by including two-body IQP generators on the graph edges:
\begin{equation}
    \mathcal{U}_{\mathrm{IQP}}^{G}(\bm{\theta})
    =
    \prod_{(j,h)\in E}
    \exp\!\left(
        -\mathrm{i}\theta_{jh} X_jX_h
    \right),
\end{equation}
possibly together with single-qubit generators or additional low-weight terms. Unlike the case of FF-PCE, which can reach maximum expressivity in a quadratic number of gates, the IQP ansatz in general needs an exponential number of gates to reach maximum expressivity (see Appendix~\ref{app:iqp_maximum_expressivity}). Thus, it is essential to have a strategy to select an appropriate subset of gates to be included in the ansatz, which we choose to be adapted to the structure of the optimisation problem. In this construction, the encoded variables remain local, while the variational circuit introduces trainable correlations along the same pairs of variables that appear in the objective or constraints.  For non-graph problems, the same principle can be applied by choosing a polynomial-size set of IQP generators that reflects the structure of the target objective, for example low-weight all-to-all interactions or interactions derived from constraint incidence patterns. In all cases, the resulting IQP-PCE model remains efficiently trainable as long as the encoded observables are restricted to $Z$-type correlations whose expectation values are classically tractable for the chosen IQP family.

We note that this graph-structured IQP construction uses a $1$-encoding, so the number of encoded variables is equal to the number of qubits.
It therefore does not provide qubit compression in the usual sense of PCE.
However, this does not affect the role of the construction in the present work, since our focus is on an efficiently simulable, quantum-inspired realisation of the PCE pipeline rather than on reducing the qubit requirements of a near-term quantum implementation.

\paragraph{Classical complexity.}
The classical cost of IQP-PCE is governed by the cost of estimating the encoded $Z$-type expectation values. Let $m=|\Pi|$ denote the number of encoded observables and let $r$ denote the number of IQP generators, or equivalently the number of ansatz parameters. For parameterised IQP circuits, Pauli-$Z$-word expectation values can be estimated to inverse-polynomial additive precision in classical polynomial time and space using the algorithm of den Nest, as employed in the IQPopt-based training framework of Ref.~\cite{recioarmengol2026trainclassicaldeployquantum}. For fixed precision, Ref.~\cite{recioarmengol2026trainclassicaldeployquantum} reports expectation-value estimation costs that scale linearly with both the number of qubits and the number of circuit parameters. Under this fixed-precision estimate, evaluating all $m$ encoded observables requires
\begin{equation}
    \mathcal{O}\!\left(m(n+r)\right)
\end{equation}
classical time per forward evaluation, up to the cost of the classical objective function. Since $m$ and $r$ are polynomial in $n$ for the IQP-PCE ansatzes considered here, the overall expectation-evaluation cost remains polynomial in the system size.

For sparse graph-structured ansatzes, $r=\mathcal{O}(n)$, giving $\mathcal{O}(mn)$ fixed-precision scaling. For dense all-to-all two-body ansatzes, $r=\mathcal{O}(n^2)$, giving $\mathcal{O}(mn^2)$ scaling. In the $1$-encoding setting, where each encoded variable is associated with a single-qubit $Z_i$ expectation value and $m\sim n$, these estimates become $\mathcal{O}(n^2)$ for sparse graph-structured ansatzes and $\mathcal{O}(n^3)$ for dense all-to-all two-body ansatzes.

\subsection{Efficiently Simulable PCE as a Dequantised Baseline}
\label{subsec:dequantised-baseline}

The free-fermionic and IQP constructions demonstrate that PCE admits nontrivial classically simulable realisations when the ansatz and encoding are chosen as compatible pairs. These realisations preserve the defining features of PCE---Pauli-observable encodings, continuous optimisation over ansatz parameters, and sign-based decoding---while replacing quantum measurement with polynomial-time classical evaluation of the required correlations.
Beyond their use as standalone heuristics, it is precisely this equivalence of structure, combined with classical tractability, that lets efficiently simulable PCE serve as a dequantised baseline. FF-PCE and IQP-PCE do not exhaust the possible classically simulable correlation models, but they show that restricted circuit families can already generate optimisation-relevant correlations within the PCE framework. Therefore, any practical advantage claimed by a quantum PCE implementation should be assessed relative to such efficiently simulable versions of the same pipeline, in addition to conventional classical optimisation baselines (a point we return to in Sec.~\ref{sec:discussion}). The same reasoning applies more broadly to expectation-value-based quantum heuristics in optimisation and machine learning, where classically simulable ansatz--encoding pairs provide a natural point of comparison.

\section{Numerical Results}
\label{sec:exp-results}

We evaluate the proposed efficiently simulable PCE variants on four families of discrete optimisation problems: MaxCut, Maximum Independent Set (MIS), the Multi-Dimensional Knapsack Problem (MDKP), and Max3SAT. Full implementation details are provided in Appendix~\ref{app:technical_details}.

\subsection{MaxCut and Maximum Independent Set on random graphs}
Let $G=(V,E)$ be an undirected graph with $|V|=n$. In MaxCut, the goal is to partition $V$ into two sets so as to maximise the number of edges crossing the partition. Equivalently, using spin variables $x_i\in\{-1,1\}$, the objective is
\[
    \max_{x\in\{-1,1\}^{n}} \frac{1}{2}\sum_{(i,j)\in E} (1-x_i x_j).
\]
In Maximum Independent Set, the goal is to find the largest subset of vertices containing no adjacent pair. Using binary variables $y_i := (x_i + 1) / 2, y_i\in\{0,1\}$, this can be written as
\[
    \max_{y\in\{0,1\}^{n}} \sum_i y_i
    \quad \text{s.t.} \quad
    y_i+y_j \leq 1 \ \ \forall (i,j)\in E.
\]

We benchmark FF-PCE and IQP-PCE on random $k$-regular graphs with $k\in\{3,4,5\}$ and up to $300$ vertices. For each pair $(n,k)$, we generate $5$ random graph instances and run each model with $5$ independent parameter initialisations (restarts). We report the approximation ratio, defined as the value of the returned solution divided by the optimal value of the corresponding instance. For MaxCut, the value is the cut size; for MIS, it is the cardinality of the feasible independent set. Reference values for MaxCut were obtained with the Burer-Monteiro algorithm~\cite{Burer2003ANP} from \texttt{pymqlib}~\cite{aboev_pymqlib}. Values for MIS were obtained with \texttt{Gurobi}~\cite{gurobi}.

\begin{figure*}[t]
    \centering
    {
    \begin{subfigure}[t]{0.8\textwidth}
        \centering
        \includegraphics[width=\textwidth]{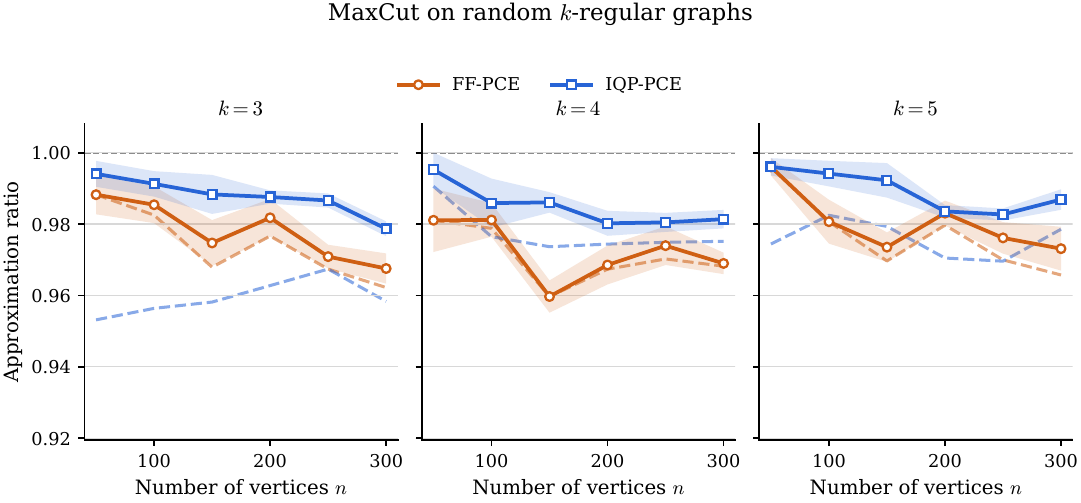}
        \caption{MaxCut}
        \label{fig:MaxCut}
    \end{subfigure}
    \hfill
    \begin{subfigure}[t]{0.8\textwidth}
        \centering
        \includegraphics[width=\textwidth]{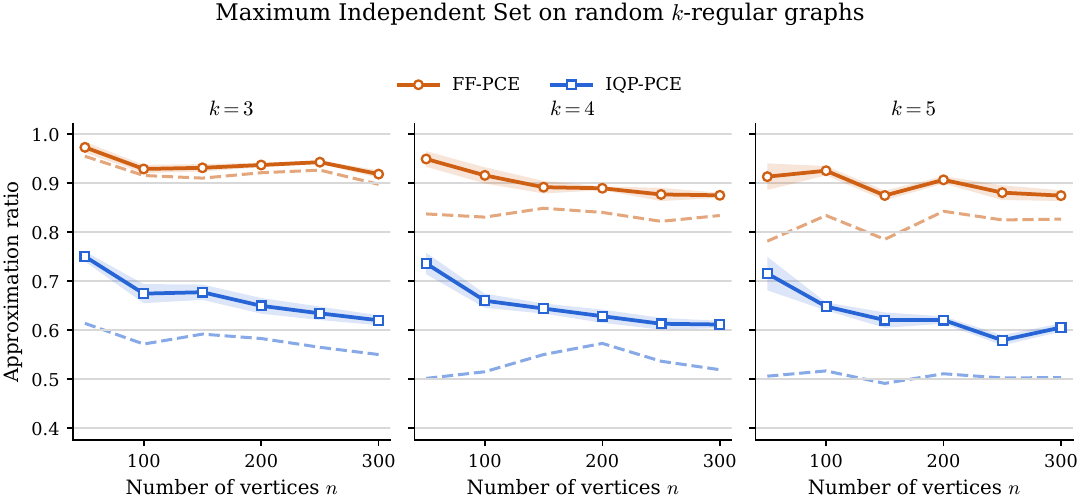}
        \caption{MIS}
        \label{fig:MIS}
    \end{subfigure}
    }
    \caption{
    Results on random $k$-regular graphs for $k\in\{3,4,5\}$.
    Panel (a) reports MaxCut approximation ratios, while panel (b) reports Maximum Independent Set approximation ratios.
    Solid curves correspond to solutions after one greedy local-search pass, whereas dashed curves show the raw decoded solutions before local search.
    FF-PCE achieves consistently strong performance on both tasks, whereas IQP-PCE attains the best MaxCut results but exhibits a marked degradation on MIS.
    }
    \label{fig:maxcut_mis_combined}
\end{figure*}

All reported approximation ratios are computed after a single greedy local-search pass applied to the decoded bitstring as in Ref.~\cite{sciorilli2024largescalequantumoptimizationsolvers}; the precise repair and local-search procedure is detailed in Appendix~\ref{app:local-search}. To separate the contribution of the PCE model from that of the classical postprocessing step, we also report the raw decoded solution quality before local search using dashed curves. Our results are shown in Fig.~\ref{fig:maxcut_mis_combined}.

For each restart, we randomly initialise the parameters and train the model with Adam~\cite{ADAM} for up to $150$ iterations. After that, we select the parameters leading to the lowest loss. The selected parameters are then used to fully train the model for up to $1500$ iterations.

\paragraph{MaxCut}
The MaxCut results are shown in Fig.~\ref{fig:MaxCut}. Across all tested graph degrees, FF-PCE achieves consistently high solution quality, with approximation ratios always above approximately $0.96$. Moreover, the raw and postprocessed curves are close to each other, indicating that FF-PCE already produces strong decoded solutions before local search and that the classical refinement provides only a modest additional improvement.

IQP-PCE achieves the best MaxCut performance overall, with approximation ratios typically above $0.98$ and often close to the optimum. In contrast to FF-PCE, the gap between the dashed and solid IQP-PCE curves is more visible, especially on $3$-regular graphs. This suggests that the IQP ansatz learns a useful global structure, but the final decoded bitstring benefits substantially from the greedy local-search pass.

As a qualitative reference point, we note that the observed empirical ratios are well above the classical Goemans--Williamson (GW) worst-case approximation guarantee~\cite{GW} of approximately $0.878$ and also above the commonly cited worst-case NP-hardness threshold of approximately $0.941$ for MaxCut~\cite{inapproximability}, even without  local search.

\paragraph{Max Independent Set}
Unlike MaxCut, MIS is a constrained optimisation problem. Therefore, we cannot directly apply the unconstrained PCE objective without accounting for the independence constraints. Following standard penalty-based approaches~\cite{andrew2014}, we optimise the unconstrained penalised objective \[ \max_{y\in\{0,1\}^{n}} \left[ \sum_i y_i - \lambda \sum_{(i,j)\in E} y_i y_j \right], \] where the second term penalises pairs of adjacent vertices selected simultaneously. In our experiments, we set $\lambda=1.5$.

Since the decoded solution may still violate some independence constraints, we apply a feasibility-repair step before evaluating the approximation ratio. Whenever the decoded set contains adjacent selected vertices, the repair heuristic iteratively removes vertices involved in constraint violations until a valid independent set is obtained. The resulting feasible solution is then improved by the same greedy local-search pass used in the MaxCut experiments.

The MIS results are shown in Fig.~\ref{fig:MIS}. FF-PCE performs robustly across all graph degrees, with postprocessed approximation ratios typically between approximately $0.88$ and $0.97$. The local-search improvement is again relatively small, suggesting that the FF-PCE decoded solutions are already close to the locally refined solutions.

IQP-PCE performs substantially worse on MIS, with postprocessed approximation ratios mostly between approximately $0.6$ and $0.75$, depending on the graph degree and size. The gap between FF-PCE and IQP-PCE is therefore much larger for MIS than for MaxCut. A plausible explanation is that the graph-structured IQP ansatz used here is well aligned with the pairwise MaxCut objective, whereas the penalised MIS landscape combines a cardinality objective with hard feasibility constraints, making the selected IQP parametrisation less effective. This difference is consistent with the greater difficulty of MIS as a constrained problem. Finally, we note that unlike MaxCut, MIS does not have a Goemans--Williamson-type approximation benchmark that can be used as a standard universal reference point, as the problem is APX-hard~\cite{APX}.

\paragraph{Comparison with linear-entangling PCE}
Training a standard PCE ansatz with gradient-based methods becomes computationally expensive at these graph sizes, since the corresponding expectation values require either quantum-hardware estimation or exponentially costly classical simulation. To obtain a feasible comparison with a conventional PCE construction, we consider a (quantum) linear-entangling PCE baseline, specified in Appendix~\ref{app:linear-entangling-pce}, and optimise all models with COBYLA~\cite{Powell1994} for $10\,000$ iterations on random $3$-regular graphs with up to $200$ vertices.

\begin{figure*}[t]
\centering

\begin{minipage}[t]{0.8\textwidth}
\centering
\vspace{0pt}
\includegraphics[width=\textwidth]{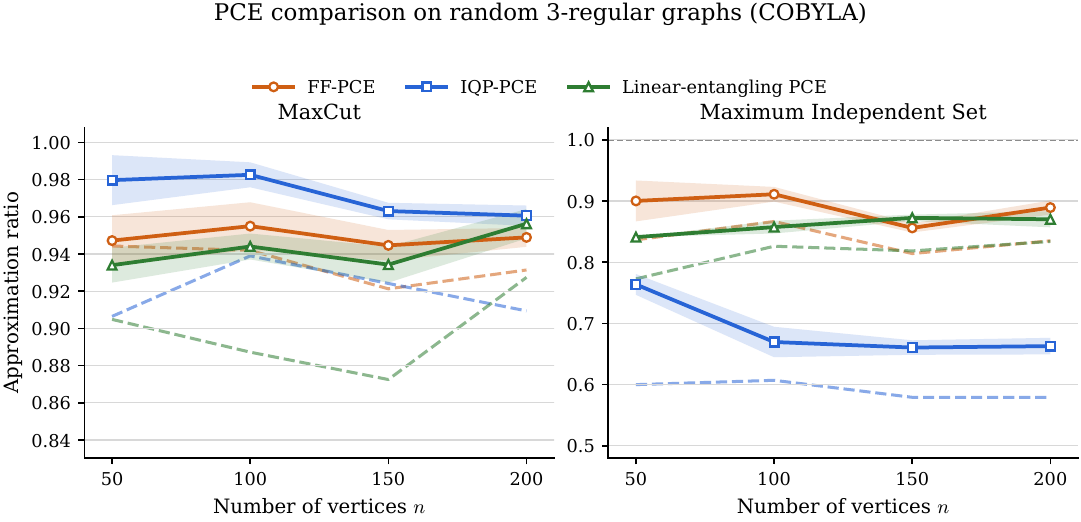}
\end{minipage}
\hfill
\begin{minipage}[t]{0.39\textwidth}
\centering
\vspace{0pt}

\captionof{table}{Resource counts for random $3$-regular graph experiments.}
\label{tab:pce-resource-counts}

\small
\sisetup{group-separator={,}}
\begin{tabular}{@{}r *{3}{S[table-format=3.0] S[table-format=4.0]}@{}}
\toprule
& \multicolumn{2}{c}{FF-PCE}
& \multicolumn{2}{c}{IQP-PCE}
& \multicolumn{2}{c}{Linear PCE} \\
\cmidrule(lr){2-3}
\cmidrule(lr){4-5}
\cmidrule(l){6-7}
$n$
& {Qubits} & {Gates}
& {Qubits} & {Gates}
& {Qubits} & {Gates} \\
\midrule
50  & 6  & 132 & 50  & 75  & 8  & 223 \\
100 & 8  & 240 & 100 & 150 & 11 & 438 \\
150 & 9  & 306 & 150 & 225 & 13 & 596 \\
200 & 11 & 462 & 200 & 300 & 15 & 822 \\
\bottomrule
\end{tabular}
\end{minipage}

\caption{
Comparison between FF-PCE, IQP-PCE, and a linear-entangling PCE baseline on random $3$-regular graphs, using COBYLA with $10\,000$ iterations for all methods.
Left: MaxCut approximation ratios.
Right: Maximum Independent Set approximation ratios.
Solid curves report the post-local-search solution quality with shadowed area representing standard deviation, while dashed curves report the raw decoded solution before local search.
}
\label{fig:pce_cobyla_comparison}

\end{figure*}

The comparison with linear-entangling PCE is shown in Fig.~\ref{fig:pce_cobyla_comparison}. On MaxCut, IQP-PCE is the strongest method under COBYLA, followed by FF-PCE and then linear-entangling PCE for most tested sizes. On MIS, FF-PCE and linear-entangling PCE achieve comparable performance, while IQP-PCE remains significantly worse. These results suggest that the advantage of a given efficiently simulable PCE family is problem-dependent: IQP-PCE is particularly effective for MaxCut, whereas FF-PCE is more robust across both MaxCut and MIS.
Importantly, we note that linear-entangling PCE does not have a better performance than classically simulable variants, suggesting that, for a fixed optimisation procedure (e.g., loss evaluation), the different models perform similarly.

\subsection{MaxCut on Large Scale Graphs}

\begin{figure*}[t]
    \centering
    \includegraphics[width=0.95\textwidth]{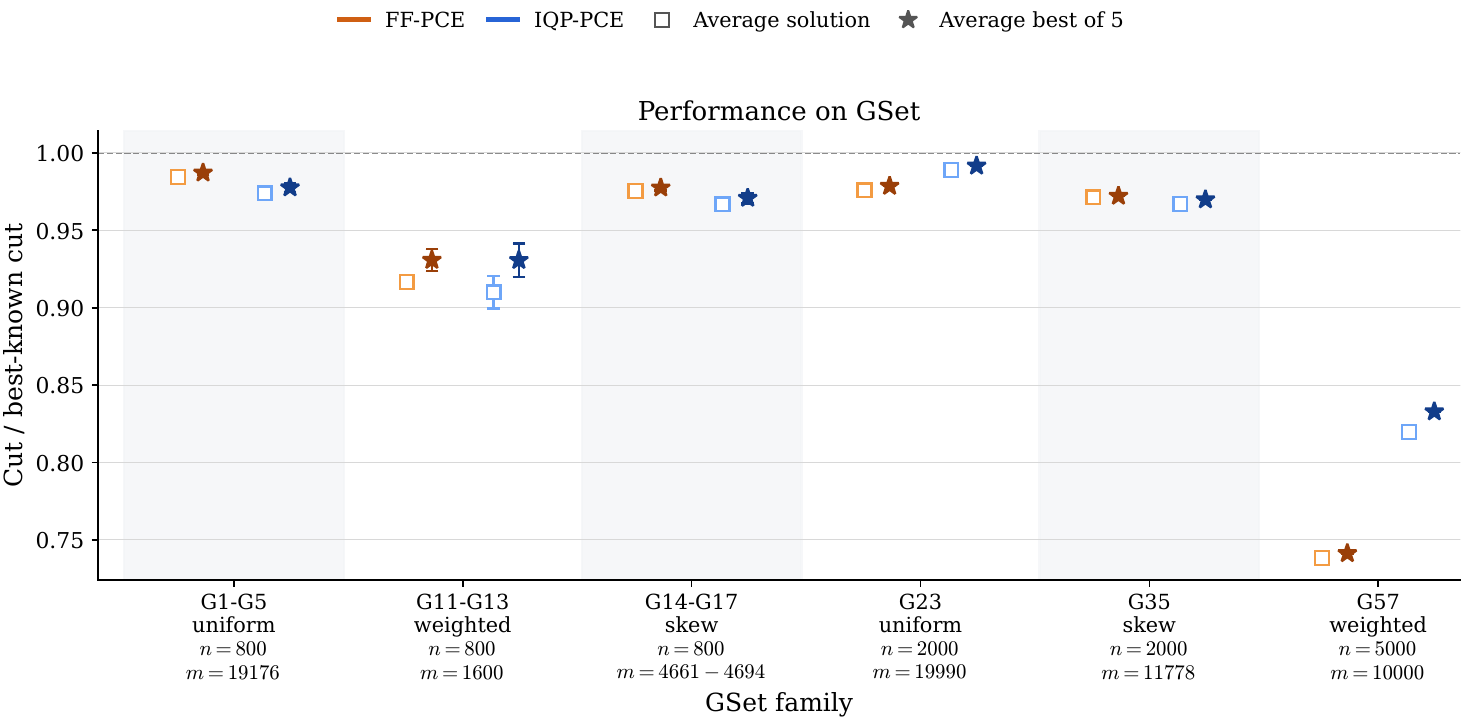}
    \caption{Results on selected instances classes from GSet, corresponding to G1-G5, G11-G13, G14-G17, G23, G35, and G57. Boxes represent the average solution over $5$ different runs and different instances; stars represent the average score of the best run for each instance. For instance families, lines represent the standard deviation between different instances. Finally, $n$ and $m$ corresponds to number of nodes and edges respectively in the corresponding class.}
    \label{fig:GSet}
\end{figure*}

To assess the performance of our approach on large-scale and complex problems, we extend our MaxCut evaluation to the well-known \texttt{Gset} dataset.
We use the same setting as the MaxCut and MIS experiments.
However, in this case the problems considered are more difficult, as they range from $800$ to $5000$ nodes, and belong to different classes, including dense, sparse, and weighted instances. For each instance, we run each model $5$ times with $5$ restarts.
Similarly to other heuristic solutions, we are interested in the average solution quality and in the approximation ratio of the best solution found.
In Fig.~\ref{fig:GSet}, we report both metrics with their standard deviation. Full numerical values are provided in Table~\ref{tab:gset-family-performance} in Appendix~\ref{app:full_results}.

We note that on most classes we obtain an approximation ratio (with regard to the best known cut value) between $0.97$ and $0.99$, highlighting the generality of our approach. Note the exception of weighted graphs, where each edge has a cost in $\{-1, 1\}$ (G11-G13, and G57). In this cases, the obtained ratios are always lower than $0.95$, consistently with the higher complexity of these instances (note that the GW ratio of $\approx 0.878$ does not hold for weighted graphs). It is interesting to note that FF-PCE and IQP-PCE obtains similar solution quality, with the exception of G57 in which IQP-PCE obtained a much higher solution quality than FF-PCE.

Finally, our approximation ratios are always comparable or higher to the ones obtained in Ref.~\cite{sciorilli2024largescalequantumoptimizationsolvers} by standard PCE with either real or simulated quantum hardware (for example, they obtained up to $0.954$ on G1, and $0.935$ on G35). This suggests that, although our models are less expressive, their easier optimisation can lead to better practical performance.

\subsection{Multi-Dimensional Knapsack Problem}

We now evaluate our models on the Multi-Dimensional Knapsack Problem (MDKP)~\cite{mdkp}, using the SAC-94 benchmark dataset~\cite{sac94_mkp}. In MDKP, we are given $n$ items, each associated with a profit $p_i$, and a set of $m$ resource constraints. Each item consumes an amount $w_{ri}$ of resource $r$, and each resource has a capacity $c_r$. The goal is to select a subset of items maximising the total profit while satisfying all capacity constraints:
\[
\max_{y\in\{0,1\}^{n}} \sum_{i=1}^{n} p_i y_i
\quad \text{s.t.} \quad
\sum_{i=1}^{n} w_{ri}y_i \leq c_r,
\quad r=1,\ldots,m.
\]

Similarly to MIS, MDKP is a constrained optimisation problem. Therefore, we optimise the following penalised unconstrained objective:
\[
-\sum_{i=1}^{n} p_i y_i
+
\lambda
\sum_{r=1}^{m}
\max\left(0,\sum_{i=1}^{n} w_{ri}y_i-c_r\right)^2
\]
with $\lambda=1.5$.

Since decoded solutions may still be infeasible, we apply a repair step before computing the approximation ratio. The repair routine greedily removes items until all capacity constraints are satisfied. At each step, it removes the selected item with the smallest profit-to-relative-weight score,
\[
\frac{p_j}{\sum_{r=1}^{m} w_{rj}/c_r}.
\]

Finally, note that since this problem is not graph-based, for the IQP ansatz we add a gate for each pair of qubits, instead of building it according to a graph structure. The reported approximation ratios are  computed using the repaired and locally improved feasible solution.

\begin{figure*}[htb]
    \centering
    \includegraphics[width=0.95\textwidth]{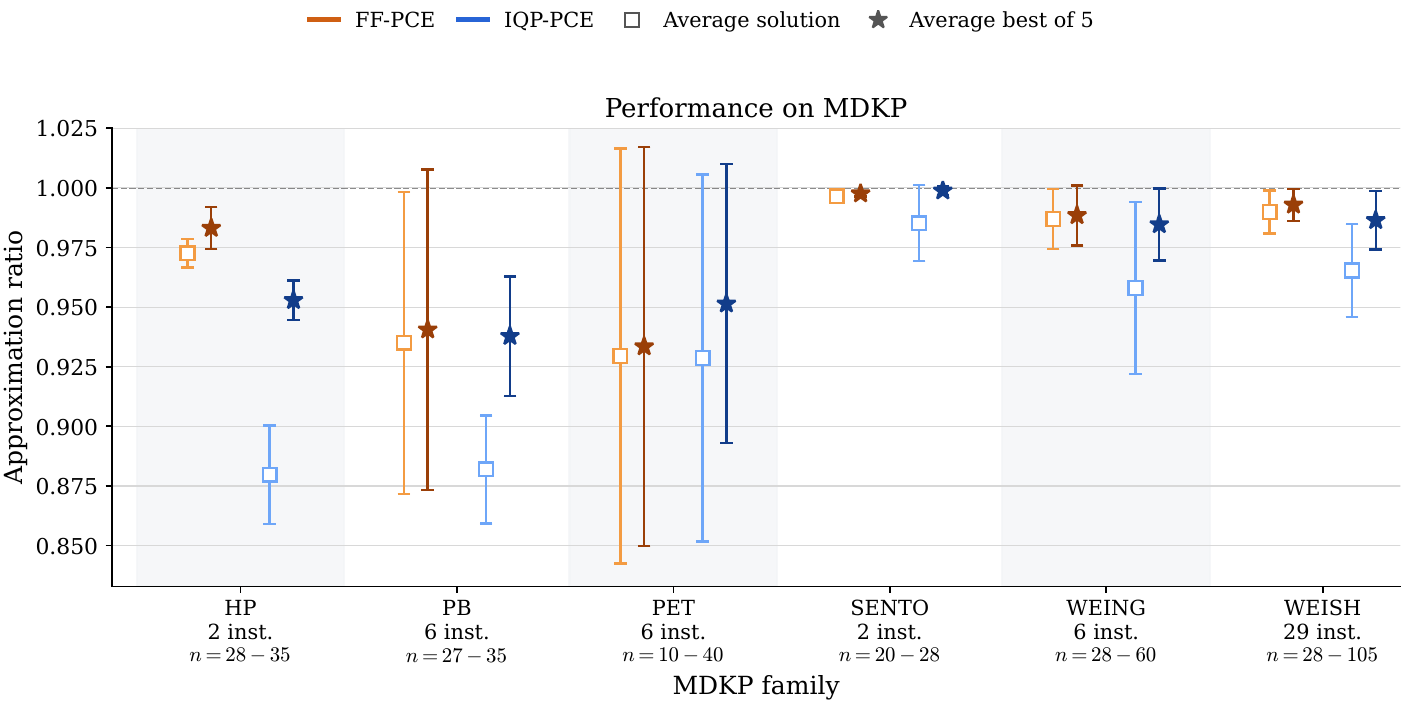}
    \caption{
        Results on the SAC-94 Multi-Dimensional Knapsack Problem benchmark.
        Instances are grouped by family: HP, PB, PET, SENTO, WEING, and WEISH.
        Empty squares report the average approximation ratio over $5$ independent runs, while stars report the average best solution found among those $5$ runs.
        Error bars indicate the standard deviation across the corresponding family.
        All reported values are computed after feasibility repair and single-bit-flip local search.
        FF-PCE achieves consistently high solution quality across most families, while IQP-PCE is more variable but remains competitive on PET, SENTO, WEING, and WEISH.
    }
    \label{fig:mdkp}
\end{figure*}

The results are shown in Fig.~\ref{fig:mdkp}. We group the SAC-94 instances by family: HP, PB, PET, SENTO, WEING, and WEISH. For each family, we report both the average approximation ratio over $5$ independent runs and the average best solution found among those $5$ runs. The approximation ratio is computed as the feasible profit of the returned solution divided by the reference value of the corresponding instance. Full numerical results, including additional information on the instances, are given in Table~\ref{tab:mdkp-family-performance-with-dimensions} in Appendix~\ref{app:full_results}.

Overall, FF-PCE achieves strong and stable performance on MDKP. On HP, SENTO, WEING, and WEISH, the average approximation ratio is close to the reference value, with family averages of $0.973$, $0.996$, $0.987$, and $0.990$, respectively. The corresponding average best-of-$5$ ratios are even higher, reaching $0.983$, $0.998$, $0.988$, and $0.993$. This indicates that FF-PCE reliably finds high-quality feasible solutions across different MDKP families, including WEISH, which contains the largest tested instances, with up to $105$ items.

The main exceptions are the PB and PET families, where FF-PCE exhibits lower average performance and substantially larger variability. In these cases, the average ratios are approximately $0.935$ and $0.929$, respectively, with best-of-$5$ averages of $0.941$ and $0.933$. This suggests that these instance classes contain more heterogeneous or harder cases for the proposed parametrisation.

These results reinforce the trend already observed on MIS: FF-PCE appears more robust on constrained optimisation problems, whereas IQP-PCE is more sensitive to the structure of the objective. This is consistent with the MaxCut and MIS experiments, where IQP-PCE was particularly effective on the pairwise unconstrained MaxCut objective but degraded on constrained selection problems. MDKP further confirms this behaviour on a non-graph benchmark with multiple resource constraints.

In summary, the MDKP experiments show that the proposed efficiently simulable PCE variants can obtain high-quality solutions also beyond graph optimisation. FF-PCE provides the most reliable performance across the benchmark families, while IQP-PCE can still reach near-optimal solutions on several families when multiple restarts are used. The gap between the average and best-of-$5$ solutions also suggests that restart strategies play an important role for constrained problems, especially for the more variable PB and PET families.

\subsection{Max3SAT}
\label{sub:Max3SAT}

Finally, we evaluate our proposed models on the Max3SAT problem. This problem is a generalisation of the Boolean satisfiability problem (SAT), and can be formulated as follows: \textit{Given a conjunctive normal form (CNF) formula where each clause has at most 3 variables, find an assignment that satisfies the largest number of clauses}. If each clause has exactly 3 variables, a Max3SAT instance can be written as \begin{equation}
    \max_{y\in\{0,1\}^n} \sum_i \phi_i(y), \text{ for } \phi_i = z_{i_1}\vee z_{i_2} \vee z_{i_3},
\end{equation}
where each $z$ is either $y_j$ or $\neg y_j$ for some $j<n$.

As this problem is an unconstrained binary optimisation problem, it can be natively formulated in the PCE framework. The corresponding target function for a $m$-clauses CNF can be written as
\begin{equation}
    f(y) = \sum_{i<m}\left((1-z_{i_1})(1-z_{i_2})(1-z_{i_3})\right),
\end{equation}
where $z_k$ is $y_k$ if it appears without negation in the corresponding clause, or $1-y_k$ otherwise.

For example, given the $2$-clauses Max3SAT instance given by \begin{equation*}
    (y_1\vee y_2 \vee y_3), (y_1\vee \neg y_4 \vee \neg y_5),
\end{equation*}
the corresponding target function is
\begin{equation*}
    f(y)=(1-y_1)(1-y_2)(1-y_3)+(1-y_1)y_4y_5.
\end{equation*}
It is easy to verify that assignments that solve both clauses (i.e., optimal solutions of the Max3SAT instance), have a corresponding function value of $0$. In general, the value of $f(y)$ corresponds to the number of unsatisfied clauses. Note that in contrast to other considered problems, whose target function can be formulated as a Quadratic Unconstrained Binary Optimization (QUBO) problem, Max3SAT contains cubic order terms. More generally, the Max-$k$-SAT problem, where each clause contains up to $k$ variables, can be formulated as a $k$-order binary problem.

To benchmark our proposed models, we consider the Max-3-SAT benchmark~\cite{pennylane_hamlib_max3sat} extracted from the library of Hamiltonians for benchmarking quantum algorithms \textit{HamLib}~\cite{sawaya2023hamlib}. The considered instances range from $10$ to $900$ variables, and are divided according to the clauses-to-variables $m/n$ ratio in $\{2, 3, 4, 5\}$. This ratio denotes the complexity of the instance, as problems with lower ratios are easier to solve optimally.

\begin{figure*}[htb]
\centering
\includegraphics[width=0.95\textwidth]{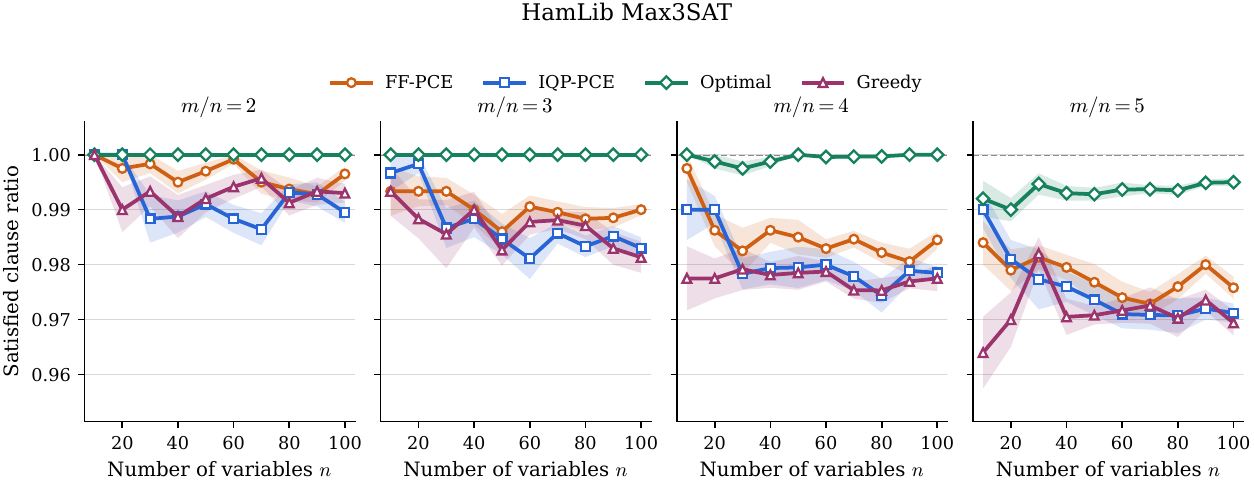}
\caption{
Results over all instances of Max3SAT benchmark~\cite{pennylaneBenchmarkingQuantum} with number of variables  $n\in\{10,\dots,100\}$. For each $n$, the dataset contains $10$ different instances per ratio $m/n$. Optimal solution is found with the RC2 algorithm~\cite{RC2_alg} implemented in the PySAT library~\cite{pysat}. Solutions for FF-PCE, IQP-PCE, and Greedy are obtained with $5$ restarts and a final local search step. Shadowed area represents the standard deviation.
}
\label{fig:max3sat_100}
\end{figure*}

As a first evaluation, we consider all instances with number of variables $n\in\{10,20,\dots,100\}$, for which is possible to obtain the optimal solution with an exact solver. In Fig.~\ref{fig:max3sat_100}, we report the solution score, computed as the fraction of clauses satisfied over the total clauses. As before, we report FF-PCE (in orange) and IQP-PCE (in blue), each with $5$ restarts and with local search at the end of the best restart. In addition, we report the optimal solution (in teal), computed with the RC2 algorithm~\cite{RC2_alg} (note that instances for $100$ variables required up to one hour of computation time). Finally, we consider a greedy solver (in purple), corresponding to selecting $5$ random assignments, and then applying the local search step to the one with the highest score.

FF-PCE obtained for each $m/n$ ratio a better satisfied clause ratio than other heuristic algorithms. On the other hand, IQP-PCE obtains a high performance up to $20$ variables, which degrades quickly when increasing the number of variables, converging towards the performances of the greedy solver. This suggests that better initial biases should be considered when constructing IQP circuits for this specific problem.

\begin{figure*}[htb]
    \centering
    \includegraphics[width=0.95\textwidth]{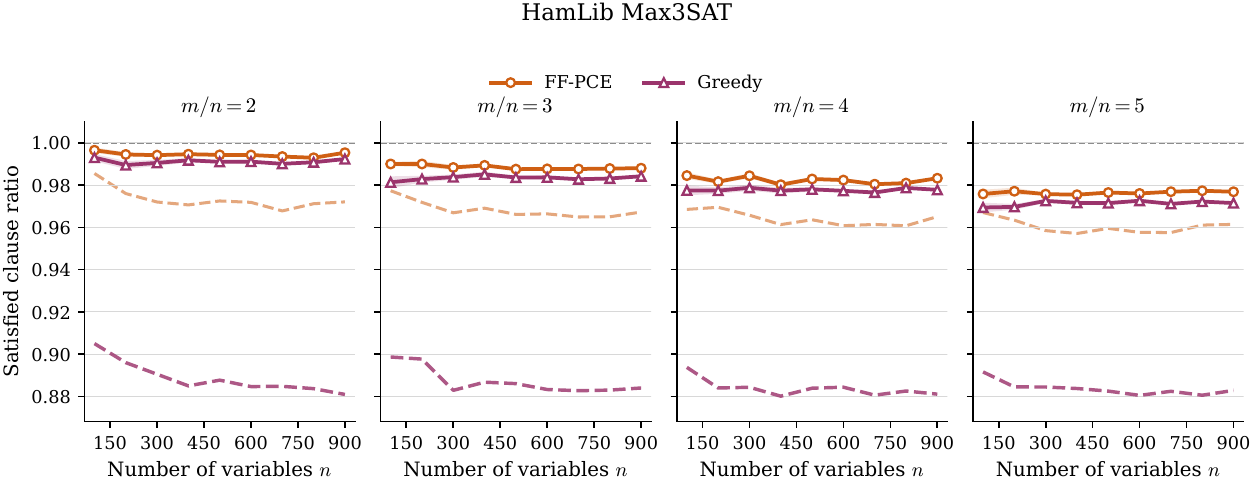}
    \caption{
        Results over all instances of Max3SAT benchmark~\cite{pennylaneBenchmarkingQuantum} with number of variables  $n\in\{100,\dots,900\}$. For each $n$, the dataset contains $10$ different instances per ratio $m/n$. Solid lines correspond to the algorithm performance after local search step, while dashed lines are the results obtained without local search (but after $5$ restarts).
    }
\label{fig:max3sat_900}
\end{figure*}

We then evaluate FF-PCE and Greedy solver on higher dimensional problems. In particular, we consider all the highest dimensional instances of HamLib, with number of variables in $\{100,\dots,900\}$. We report our results in Fig.~\ref{fig:max3sat_900}, this time also with the solutions obtained without local search. Full numerical results for FF-PCE are reported in Table~\ref{tab:max3sat-ffpce-compact}. We observe that FF-PCE performance does not degrade with increasing sizes. In addition, the raw solutions found without local search have always a satisfied clause ratio of at least $0.96$, suggesting that the algorithm is able to find consistently good results. On the other hand, the greedy raw solution, even with $5$ restarts, has a ratio converging to the $7/8=0.875$ value, corresponding to the expected approximation ratio of a random initial assignment. Note that under $P\neq NP$, this threshold is optimal for polynomial-time approximation of Max3SAT (where each clauses has exactly three literals): for every $\varepsilon>0$, no polynomial-time algorithm can guarantee an approximation ratio better than $7/8+\varepsilon$~\cite{inapproximability}.

\section{Discussion and Outlook}
\label{sec:discussion}

In this work, we introduced \emph{efficiently simulable Pauli Correlation Encoding} as a quantum-inspired, fully classical realisation of the PCE framework. By choosing compatible pairs of ansatz families and encoded observables, the Pauli expectation values required for training and decoding can be evaluated efficiently on classical hardware. We instantiated this idea using free-fermionic and IQP circuit families, obtaining PCE variants that preserve the encoding, optimisation, and sign-based decoding structure of standard PCE while avoiding quantum measurement overhead.

Our experiments show that these variants can produce high-quality solutions on standard combinatorial optimisation benchmarks, including MaxCut, Maximum Independent Set (MIS), Multi-Dimensional Knapsack (MDKP), and Max3SAT, with an additional low-autocorrelation binary sequence (LABS) study reported in Appendix~\ref{app:LABS}. These results indicate that efficiently simulable PCE is best viewed as a family of correlation-based heuristics rather than as a single algorithm: its performance depends on the compatibility between the ansatz-induced correlation structure and the target optimisation problem. IQP-PCE is particularly well aligned with pairwise graph objectives such as MaxCut, where the ansatz connectivity can directly mirror the graph structure, whereas constrained selection problems such as MIS and MDKP require balancing objective value against feasibility, penalties, and repair, and the same IQP parametrisation is less naturally matched to this landscape. FF-PCE behaves more robustly across these tasks, suggesting that covariance-matrix correlations provide a useful generic inductive bias for PCE-style optimisation even when the problem is not explicitly fermionic. This supports viewing PCE more broadly as a correlation-based optimisation framework in which different ansatz families supply different inductive biases, and suggests using FF-PCE as a natural first choice when no problem-specific circuit structure is available, while treating IQP-PCE as a more specialised option for graph-structured correlation models. We emphasise, however, that this is an empirical guideline rather than a general dominance claim.

The role of classical postprocessing further clarifies how PCE should be interpreted as a heuristic. The decoded correlations need not by themselves produce a final locally optimal solution; rather, they provide a structured proposal that is subsequently refined by repair and local search. From this perspective, PCE is best understood as a correlation-guided initialisation or proposal mechanism that can be combined with classical improvement routines. This is consistent with the broader hybrid nature of the framework: the quantum or quantum-inspired component generates correlations, while classical computation performs optimisation, decoding, feasibility repair, and refinement.

A central implication of the results is the distinction between formal expressivity and practical optimisability. A universal or more general quantum ansatz may represent a larger class of correlations, but this additional expressivity is useful only if it can be accessed within the available optimisation and measurement budget. The comparison with a linear-entangling PCE baseline illustrates this point: under the tested optimisation protocol, the efficiently simulable variants are competitive with, and in some cases outperform, a standard quantum PCE ansatz. Practical performance for PCE is therefore controlled not only by the expressivity of the ansatz, but also by the cost and stability of evaluating and optimising the encoded correlations.

The main contribution of this work is methodological: efficiently simulable PCE provides a dequantised baseline for assessing future quantum implementations of PCE. Our results do not rule out useful quantum implementations, but they clarify what such implementations must demonstrate. A quantum ansatz should provide optimisation gains that cannot already be reproduced by efficiently simulable ansatz--encoding pairs, and those gains must be large enough to compensate for the cost of estimating expectation values on quantum hardware. Benchmarking quantum PCE only against conventional classical optimisation algorithms is therefore insufficient; it should also be compared against classically simulable realisations of the same PCE pipeline.

This reframes the search for quantum utility in PCE. The relevant question is not merely whether a quantum circuit can generate correlations that are hard to simulate in general, but whether the specific correlations useful for a given optimisation problem are both practically valuable and inaccessible to efficient classical correlation models. Efficiently simulable PCE provides a way to test this distinction empirically: problem classes where hardware-native or universal ansatz consistently outperform these dequantised baselines would constitute the stronger candidates for practical quantum PCE utility.

Several directions remain open. Efficiently simulable PCE could serve as a classical warm-start for quantum PCE solvers, training an FF-PCE or IQP-PCE model classically and using its optimised parameters to initialise a more expressive quantum ansatz, for instance by adding non-simulable layers on top of the simulable circuit. Starting the quantum optimisation from such a configuration, rather than from a random one, might reduce the number of circuit evaluations needed and let the quantum model refine correlations that the simulable ansatz cannot represent on its own; we have not tested this here, and leave a proper evaluation of the idea, and of when it pays off, to future work. The class of efficiently simulable PCE models could also be extended, for example to Clifford circuits, tensor-network ansatzes, and problem-specific graphical models, which may provide even stronger baselines, and the optimisation, hyperparameter, restart, repair, and local-search strategies used here are not exhaustive. Finally, it will be important to develop resource-aware comparisons between quantum and dequantised PCE, including optimisation cost, measurement cost, postprocessing, and time-to-solution, which may help identify problem classes where genuinely quantum correlation models provide a practical advantage within the PCE framework.

\begin{acknowledgments}
We thank Matthew DeCross and Fr\'ed\'eric Sauvage for their careful review of the manuscript.
\end{acknowledgments}

\bibliography{refs.bib}

\clearpage
\onecolumngrid
\appendix

\section{Technical Details}
\label{app:technical_details}
This appendix summarises the implementation details used in the experiments.
All models follow the same high-level PCE pipeline, based on the information provided in \cite{sciorilli2024largescalequantumoptimizationsolvers}.

\subsection{Hyperparameters selection}

In each experiment, the target loss is of the form
\begin{equation}
    \mathcal{L}(\theta) = f(\tanh(\alpha \langle \Pi \rangle)) + \beta \mathcal{L}^{(\text{reg})},
\end{equation}
where $f$ in the target function (including the penalty component in constrained problems), and  $\mathcal{L}^{(\text{reg})}$ is the regularisation loss obtained from  $\left(
        \frac{1}{m}\sum_{i=1}^{m}s_i^2
    \right)^2 .$ Since $s_i^2$ is small when the relaxed variable is close to zero and large
when it is close to a binary sign, this term penalises overly saturated
relaxed assignments during training.  Therefore $\mathcal{L}^{(\text{reg})}$ acts as smoothness/scale
regulariser for the continuous PCE relaxation.

The hyperparameters $\alpha$ and $\beta$ play a major role in the effectiveness of the optimisation process. In all FF-PCE and IQP-PCE experiments reported, the loss-rescaling
parameters for a $m$ variables problem are
\begin{equation}
    \alpha = \sqrt{m},
    \qquad
    \beta = \frac{m}{2}.
\end{equation}
These values were obtained from a gridsearch on $\alpha_0, \beta_0, k_\alpha, k_\beta$ $\alpha(m)=\alpha_0 m^{k_\alpha}$,  $\beta(m)=\beta m^{k_\beta}$. Scaling factors $\alpha_0$ and $\beta_0$ were selected from $[0.25, 0.5, 1, 2]$, while exponents $k_\alpha$ and $k_\beta$ were selected from $0.25, 0.5, 1, 1.5$. These selections were based on the analysis provided in \cite{sciorilli2024largescalequantumoptimizationsolvers}, in which it is suggested that optimal $\alpha$ and $\beta$ parameters scale polynomially with problem size $m$. To select the optimal values, we used as target function the average raw approximation ratio over 4 random MaxCut graphs of size from $50$ to $200$. We expect better hyperparameters explorations to obtain better results, in particular on different problems.

On the other hand, for the linear entangling PCE on MaxCut we used $\alpha = 1.5m, \beta = 0.5m$, as suggested in the original paper.

\subsection{Models implementation}
\paragraph{FF-PCE}

The matchgate ansatz consists of alternating layers of single-qubit
$R_Z$ rotations and nearest-neighbour $R_{XX}$ rotations. In the implemented
solver, the number of layers is fixed to
\begin{equation}
    L_{\mathrm{FF}} = 2 n_{\mathrm{FF}} .
\end{equation}
Each layer contains $n_{\mathrm{FF}}$ parameters for the $R_Z$ rotations and
$n_{\mathrm{FF}}-1$ parameters for the $R_{XX}$ rotations. Hence the total
number of FF-PCE trainable parameters is
\begin{equation}
    N_{\mathrm{par}}^{\mathrm{FF}}
    =
    L_{\mathrm{FF}}\,(2n_{\mathrm{FF}}-1).
\end{equation}
The FF-PCE implementation is written in PyTorch and optimises the covariance
matrix objective directly. Parameters are randomly initialised from a normal
distribution. By default the solver performs five short warm-up trials, each
lasting one tenth of the requested optimisation budget, selects the trial with
lowest warm-up loss, and then runs the full optimisation from that point.

\paragraph{IQP-PCE}
The IQP-PCE solver uses commuting rotations generated by tensor products of
Pauli $X$ operators and estimates expectation values of Pauli-$Z$ observables.
The implementation maps variable $i$ to the $i$-th qubit, i.e. $n_{\mathrm{IQP}}=m$ qubits for the IQP solver.

For graph problems, the IQP ansatz is constructed from the graph topology:
for each edge $(i,j)$, we add one commuting IQP gate on qubits assigned to variables $i$ and $j$. With
weight-one encodings this is a two-qubit gate on the corresponding pair of
vertices. For MDKP, we select a gate for each pair of qubits.

The IQP solver is implemented using JAX together with the
\texttt{iqpopt} \cite{iqpopt} expectation-value routines, estimated with $1000$ shots. As for the FF-PCE,  the optimisation process performs
five random trials for  one tenth of the total optimisation budget, and keeps the parameters with the lowest final training
loss.

\paragraph{Linear-entangling PCE baseline}
\label{app:linear-entangling-pce}
For the comparison against a standard PCE ansatz, we use a CUDA-Q \cite{CudaQ}
implementation with an linear connectivity circuit: each repetition
contains single-qubit $R_Y$ and $R_Z$ rotations on every qubit followed by a
linear chain of CNOT gates, plus a final rotation layer. For the encoding, we use the {original} PCE mapping with weight $w=2$.
This mapping assigns variables to homogeneous Pauli strings supported on
$w$ qubits: approximately one third of the variables are encoded with
$X$-type strings, one third with $Y$-type strings, and the remainder with
$Z$-type strings. This gives a quadratic reduction in the number of variables.

For the number of qubits, we used $n=\sqrt{m}+1$ and $1.2n$ layers. This follows the scaling used in the original PCE implementation. Note that while using more layers may increase expressivity, it can also increase the training complexity, possibly leading to a lower final performance (fixed the shot budget).

Finally, the linear-entangling PCE scripts use the original PCE hyperparameter scaling
\begin{equation}
    \alpha
    =
    1.5\,m^{\lfloor w/2\rfloor},
    \qquad
    \beta
    =
    \frac{d+1}{8}\,m ,
\end{equation}
where $d$ is the regular degree of the graph. In the $3$-regular comparison
experiments, with weight $w=2$, this gives
\begin{equation}
    \alpha=1.5m, \qquad \beta=\frac{m}{2}.
\end{equation}

\subsection{Decoding, repair, and local search.}
\label{app:local-search}
All reported solutions are obtained from the decoded expectation signs. For
MaxCut, the decoded partition is scored directly by summing the weights of
cut edges. For MIS, the selected set is repaired before scoring: while the
set contains an edge with both endpoints selected, one endpoint of the
violating edge is removed at random, and the scan is restarted. The repaired
independent set size is then used as the feasible score.

For MDKP, the selected item set is repaired greedily. While any capacity is
violated, the code removes the selected item with the smallest
profit-to-relative-weight score
\begin{equation}
    \frac{p_j}
    {\sum_{r=1}^{d} w_{rj}/(c_r+10^{-9})}.
\end{equation}
The final MDKP score is the total profit of the repaired feasible set.

We then apply a greedy single-bit-flip local search. The routine sweeps over all
variables, accepts any flip that strictly improves the repaired feasible score,
and repeats these sweeps until no improving flip is found.

\section{Full Numerical Results}
\label{app:full_results}

This appendix reports the complete numerical results for our evaluation MaxCut on GSet, and MDKP on SAC-94 \cite{sac94_mkp}. In particular, we report in Table~\ref{tab:gset-family-performance} the full results on the GSet benchmark instances, and in Table~\ref{tab:mdkp-family-performance-with-dimensions} the average results per class instance on the SAC-94 dataset.

\begin{table}[t]
\centering
\caption{Performance on GSet. For each instance, we report the number of nodes $n$, number of edges $m$, and for each model the average solution quality and the best solution found in $5$ runs.}
\label{tab:gset-family-performance}
\begin{tabular}{@{}llrr *{2}{cc}@{}}
\toprule
& & & & \multicolumn{2}{c}{FF-PCE} & \multicolumn{2}{c}{IQP-PCE} \\
\cmidrule(lr){5-6}\cmidrule(l){7-8}
Instance & Type & $n$ & $m$ & Avg. solution & Avg. best of 5 & Avg. solution & Avg. best of 5 \\
\midrule
G1 & uniform & 800 & 19176 & 0.984 & 0.985 & 0.973 & 0.975 \\
G2 & uniform & 800 & 19176 & 0.986 & 0.989 & 0.976 & 0.981 \\
G3 & uniform & 800 & 19176 & 0.984 & 0.987 & 0.975 & 0.979 \\
G4 & uniform & 800 & 19176 & 0.985 & 0.987 & 0.972 & 0.975 \\
G5 & uniform & 800 & 19176 & 0.984 & 0.987 & 0.974 & 0.976 \\
G11 & weighted & 800 & 1600 & 0.913 & 0.926 & 0.898 & 0.918 \\
G12 & weighted & 800 & 1600 & 0.918 & 0.939 & 0.916 & 0.939 \\
G13 & weighted & 800 & 1600 & 0.919 & 0.928 & 0.916 & 0.935 \\
G14 & skew & 800 & 4694 & 0.977 & 0.979 & 0.972 & 0.976 \\
G15 & skew & 800 & 4661 & 0.975 & 0.976 & 0.964 & 0.970 \\
G16 & skew & 800 & 4672 & 0.976 & 0.979 & 0.966 & 0.967 \\
G17 & skew & 800 & 4667 & 0.974 & 0.976 & 0.966 & 0.970 \\
G23 & uniform & 2000 & 19990 & 0.976 & 0.979 & 0.989 & 0.992 \\
G35 & skew & 2000 & 11778 & 0.971 & 0.972 & 0.967 & 0.970 \\
G57 & weighted & 5000 & 10000 & 0.738 & 0.741 & 0.820 & 0.833 \\
\bottomrule
\end{tabular}
\end{table}

\begin{table}[t]
\centering
\caption{Performance on MDKP families. Values are mean $\pm$ standard deviation across instances in each family. The $n$ column corresponds to the ranges of problem dimension (i.e., the number of variables/items in the knapsack); the dimension $m$ is the number of resources constraints.}
\label{tab:mdkp-family-performance-with-dimensions}
\begin{tabular}{@{}lrrr *{2}{cc}@{}}
\toprule
& & & & \multicolumn{2}{c}{FF-PCE} & \multicolumn{2}{c}{IQP-PCE} \\
\cmidrule(lr){5-6}\cmidrule(l){7-8}
Class & Inst. & $n$ & $m$ & Avg. solution & Avg. best of 5 & Avg. solution & Avg. best of 5 \\
\midrule
HP & 2 & 28-35 & 4 & 0.973 $\pm$ 0.006 & 0.983 $\pm$ 0.009 & 0.880 $\pm$ 0.021 & 0.953 $\pm$ 0.008 \\
PB & 6 & 27-35 & 2-4 & 0.935 $\pm$ 0.063 & 0.941 $\pm$ 0.067 & 0.882 $\pm$ 0.023 & 0.938 $\pm$ 0.025 \\
PET & 6 & 10-40 & 10-30 & 0.929 $\pm$ 0.087 & 0.933 $\pm$ 0.084 & 0.929 $\pm$ 0.077 & 0.951 $\pm$ 0.059 \\
SENTO & 2 & 20-28 & 10 & 0.996 $\pm$ 0.000 & 0.998 $\pm$ 0.001 & 0.985 $\pm$ 0.016 & 0.999 $\pm$ 0.002 \\
WEING & 6 & 28-60 & 2-30 & 0.987 $\pm$ 0.013 & 0.988 $\pm$ 0.013 & 0.958 $\pm$ 0.036 & 0.985 $\pm$ 0.015 \\
WEISH & 29 & 28-105 & 2-5 & 0.990 $\pm$ 0.009 & 0.993 $\pm$ 0.007 & 0.965 $\pm$ 0.019 & 0.986 $\pm$ 0.012 \\
\bottomrule
\end{tabular}
\end{table}

Note that \cite{PCE_for_MDKP} applies standard PCE to selected instances from HP, PB, and PET classes, obtaining approximation ratios between $0.7$ and $0.8$. However, as their constraint formulation and postprocessing are different from the ones used in this work, it is not possible to provide a fair comparison with our results.

Finally, we report full numerical results for FF-PCE over instances of the Max3SAT benchmark from HamLib \cite{sawaya2023hamlib} in Table~\ref{tab:max3sat-ffpce-compact}.

\begin{table*}[t]
\centering
\caption{FF-PCE satisfied-clause ratios on random HamLib Max3SAT. Each entry is the mean over ten instances.}
\label{tab:max3sat-ffpce-compact}
\setlength{\tabcolsep}{5.5pt}
\renewcommand{\arraystretch}{1.08}
\begin{tabular}{@{}r *{4}{cc}@{}}
\toprule
& \multicolumn{2}{c}{$m/n=2$} & \multicolumn{2}{c}{$m/n=3$} & \multicolumn{2}{c}{$m/n=4$} & \multicolumn{2}{c}{$m/n=5$} \\
\cmidrule(lr){2-3}\cmidrule(lr){4-5}\cmidrule(lr){6-7}\cmidrule(l){8-9}
Variables ($n$) & Before LS & After LS & Before LS & After LS & Before LS & After LS & Before LS & After LS \\
\midrule
10 & 1.0000 & 1.0000 & 0.9933 & 0.9933 & 0.9950 & 0.9975 & 0.9840 & 0.9840 \\
20 & 0.9825 & 0.9975 & 0.9900 & 0.9933 & 0.9838 & 0.9863 & 0.9730 & 0.9790 \\
30 & 0.9917 & 0.9983 & 0.9867 & 0.9933 & 0.9733 & 0.9825 & 0.9747 & 0.9813 \\
40 & 0.9900 & 0.9950 & 0.9808 & 0.9900 & 0.9744 & 0.9863 & 0.9655 & 0.9795 \\
50 & 0.9780 & 0.9970 & 0.9687 & 0.9860 & 0.9715 & 0.9850 & 0.9688 & 0.9768 \\
60 & 0.9867 & 0.9992 & 0.9817 & 0.9906 & 0.9729 & 0.9829 & 0.9640 & 0.9740 \\
70 & 0.9821 & 0.9950 & 0.9762 & 0.9895 & 0.9761 & 0.9846 & 0.9654 & 0.9729 \\
80 & 0.9794 & 0.9938 & 0.9746 & 0.9883 & 0.9681 & 0.9822 & 0.9645 & 0.9760 \\
90 & 0.9806 & 0.9928 & 0.9707 & 0.9885 & 0.9639 & 0.9806 & 0.9649 & 0.9800 \\
100 & 0.9855 & 0.9965 & 0.9773 & 0.9900 & 0.9685 & 0.9845 & 0.9670 & 0.9758 \\
200 & 0.9760 & 0.9945 & 0.9718 & 0.9900 & 0.9695 & 0.9816 & 0.9634 & 0.9771 \\
300 & 0.9720 & 0.9942 & 0.9669 & 0.9883 & 0.9657 & 0.9844 & 0.9585 & 0.9757 \\
400 & 0.9706 & 0.9946 & 0.9691 & 0.9893 & 0.9613 & 0.9802 & 0.9571 & 0.9754 \\
500 & 0.9725 & 0.9943 & 0.9661 & 0.9875 & 0.9636 & 0.9829 & 0.9595 & 0.9765 \\
600 & 0.9718 & 0.9943 & 0.9664 & 0.9877 & 0.9608 & 0.9824 & 0.9576 & 0.9760 \\
700 & 0.9678 & 0.9935 & 0.9649 & 0.9876 & 0.9614 & 0.9804 & 0.9575 & 0.9769 \\
800 & 0.9712 & 0.9929 & 0.9650 & 0.9878 & 0.9607 & 0.9809 & 0.9611 & 0.9773 \\
900 & 0.9721 & 0.9953 & 0.9673 & 0.9880 & 0.9651 & 0.9832 & 0.9614 & 0.9768 \\
\bottomrule
\end{tabular}
\end{table*}

\section{Wall-clock estimate for quantum PCE}
\label{app:wall-clock}
The shot-complexity estimate in Sec.~\ref{sec:PCE} can be converted into a crude wall-clock runtime by assigning an effective time \(t_{\mathrm{shot}}\) to one circuit execution and measurement. For a circuit of depth \(D\), hardware layer time \(t_{\mathrm{layer}}\), and readout/reset time \(t_{\mathrm{read}}\), we estimate
\begin{equation}
    t_{\mathrm{shot}}
    \approx
    D t_{\mathrm{layer}} + t_{\mathrm{read}} .
\end{equation}
Therefore,
\begin{equation}
    T_{\mathrm{train}}
    \sim
    N_{\mathrm{shots}}^{(\mathrm{train})}
    \left(
        D t_{\mathrm{layer}} + t_{\mathrm{read}}
    \right).
\end{equation}
Combining this expression with Eq.~\eqref{eq:grouped-shot-cost} gives
\begin{equation}
T_{\mathrm{train}}
=
\mathcal{O}\!\left[
N_t(c+2N_p)
\frac{G}{\epsilon^2}
\log\!\left(\frac{m}{\delta}\right)
\left(Dt_{\mathrm{layer}}+t_{\mathrm{read}}\right)
\right].
\end{equation}
For the concrete estimate quoted in the main text, consider $m=10^3$ encoded variables, $N_p=10^2$ trainable parameters, $N_t=10^3$ optimisation steps, and $c=1$. We take $G=3$, as in the homogeneous encoding of Ref.~\cite{sciorilli2024largescalequantumoptimizationsolvers}, and require an overall failure probability $\delta=10^{-2}$ for each loss or shifted-loss evaluation. For Pauli measurement outcomes in $\{-1,1\}$, Hoeffding's inequality and a union bound give the explicit sufficient shot count~\cite{hoeffding1963probability, majsak2025jointmeasurement}
\begin{equation}
S(\epsilon,\delta)
=
\frac{2}{\epsilon^2}
\log\!\left(\frac{2m}{\delta}\right)
\end{equation}
per measurement setting. Let $\gamma_{\min}$ be the minimum correlation magnitude over all encoded signs and all loss and shifted-loss evaluations, as defined in the main text. We conservatively choose $\epsilon=\gamma_{\min}/2$ and use this same precision throughout training. The resulting training budget is
\begin{equation}
N_{\mathrm{shots}}^{(\mathrm{train})}
\approx
N_t(c+2N_p)G S(\gamma_{\min}/2,\delta).
\end{equation}
This gives the following illustrative estimates:
\begin{center}
\begin{tabular}{@{}cccc@{}}
\toprule
$\gamma_{\min}$ & $N_{\mathrm{shots}}^{(\mathrm{train})}$
& $T$ at $40\,\mu\mathrm{s}$ & $T$ at $1\,\mathrm{ms}$ \\
\midrule
$10^{-1}$ & $5.9\times10^{9}$  & $2.7$ days & $68$ days \\
$10^{-2}$ & $5.9\times10^{11}$ & $273$ days & $19$ years \\
\bottomrule
\end{tabular}
\end{center}
These runtimes assume continuous serial circuit execution and are deliberately conservative because training need not resolve every sign at every iteration: looser precision early in training, adaptive shot allocation, gradient-free optimisation (for example COBYLA in Ref.~\cite{sciorilli2024largescalequantumoptimizationsolvers}), or high-confidence sign recovery only at final decoding can substantially reduce the cost. Nevertheless, the estimates illustrate the inverse-square dependence on the correlation margin: decreasing the relevant margin from $10^{-1}$ to $10^{-2}$ increases the shot count and runtime by a factor of one hundred. By contrast, when the observables share a fixed number $G$ of measurement settings--in particular, $G=3$ for the homogeneous encoding of Ref.~\cite{sciorilli2024largescalequantumoptimizationsolvers}--the number of observables affects the idealised bound only through the logarithmic simultaneous-confidence factor. Actual runtimes additionally depend on hardware architecture, compilation, parallelisation, readout, routing, and error correction, so these values should be interpreted as scaling estimates rather than device-level benchmarks.

\section{PCE for approximate LABS}
\label{app:LABS}

The low-autocorrelation binary-sequence (LABS) problem asks for a sequence
$s=(s_1,\ldots,s_N)$, with $s_i\in\{-1,+1\}$, that minimises the squared
aperiodic autocorrelation energy
\begin{align}
    C_k(s) &= \sum_{i=1}^{N-k} s_i s_{i+k},
    \qquad k=1,\ldots,N-1, \\
    E(s) &= \sum_{k=1}^{N-1} C_k(s)^2 .
    \label{eq:labs-energy}
\end{align}
Solution quality is usually reported through the merit factor
\begin{equation}
    F(s)=\frac{N^2}{2E(s)} ,
    \label{eq:labs-merit-factor}
\end{equation}
which is maximised by minimising $E(s)$. Since the exact optimum energies
$E_N^\star$ are known for all sizes considered here, we report the
normalised merit factor
\begin{equation}
    \rho_F(s)
    =
    \frac{F(s)}{F_N^\star}
    =
    \frac{E_N^\star}{E(s)} .
    \label{eq:labs-normalised-merit-factor}
\end{equation}
Thus, $\rho_F=1$ corresponds to an optimal LABS sequence.

LABS is often studied as an exact optimisation benchmark, where the relevant
quantity is the time required to sample or reach an optimal sequence \cite{QAOA_x_LABS}. Our goal
here is different. We use LABS as an additional approximate-optimisation test:
given a fixed training budget, we ask whether the correlations generated by
PCE decode to sequences with nontrivial merit factor. The results should
therefore not be interpreted as a Time-To-Solution (TTS) comparison with exact LABS
solvers.

Following the approach in \cite{sciorilli2026competitivenisqqubitefficientsolver},  for each encoded observable $P_i$, we define the relaxed spin
\begin{equation}
    \widetilde{s}_i(\theta)
    =
    \tanh\!\left(\alpha \langle P_i\rangle_\theta\right).
\end{equation}
The LABS training loss is obtained by substituting these relaxed variables
into Eq.~\eqref{eq:labs-energy},
\begin{equation}
    \mathcal{L}_{\mathrm{LABS}}(\theta)
    =
    \sum_{k=1}^{N-1}
    \left(
        \sum_{i=1}^{N-k}
        \widetilde{s}_i(\theta)
        \widetilde{s}_{i+k}(\theta)
    \right)^2
    -
    \beta
    \sum_{i=1}^{N}
    \widetilde{s}_i(\theta)^2 .
    \label{eq:labs-relaxed-loss}
\end{equation}

The second term discourages the relaxed variables from remaining close to
zero and favours assignments with larger spin magnitude. As in the main
experiments, we use the problem-independent scaling
$\alpha=\sqrt{N}$ and $\beta=N/2$ for each efficiently simulable PCE, without LABS-specific tuning.

We also evaluate a quantum PCE for $N$ up to $30$. The core implementation is as described in Appendix~\ref{app:technical_details}; however, based on the findings in \cite{sciorilli2026competitivenisqqubitefficientsolver}, we perform the following modifications:
\begin{itemize}
    \item First, we select the $2$-encoding maximising non-commutative pairs (NC encoding in the original paper);
    \item Second, we selected $\alpha=1.5N$ and $\beta=15$. These values were selected by maximising the average approximation ratio of the solution quality, and are therefore coherent with our setting.
\end{itemize}

\begin{figure*}[t]
    \centering
    \includegraphics[width=0.95\textwidth]{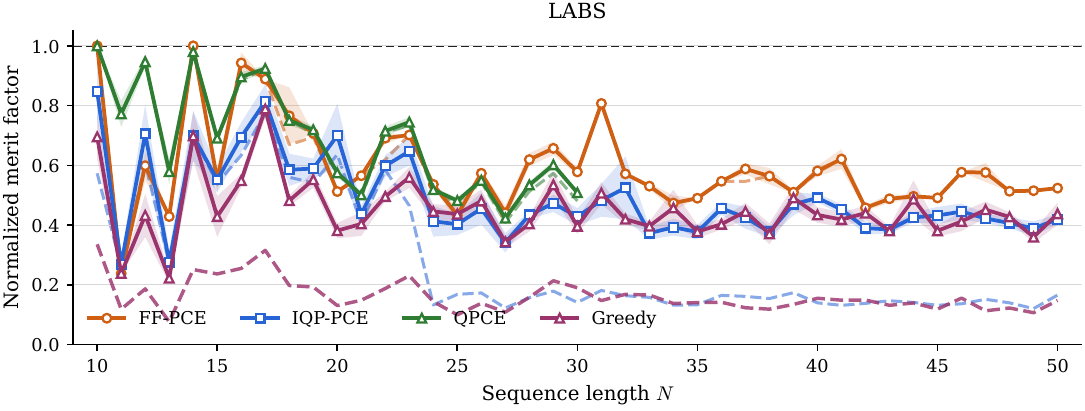}
    \caption{
    Approximate LABS performance measured by the normalised merit factor
    $\rho_F=F/F_N^\star$. Dashed curves denote raw sequences decoded from the
    signs of the PCE expectation values, while solid curves denote the same
    sequences after greedy single-spin local search. Shaded regions indicate
    one standard error of the mean for the postprocessed results, and the
    horizontal line marks the optimum $\rho_F=1$. FF-PCE and IQP-PCE use five
    seeds for every $N=10,\ldots,50$.
    }
    \label{fig:labs-approximate}
\end{figure*}

The results are shown in Fig.~\ref{fig:labs-approximate}. Since the optimal
LABS merit factors vary irregularly with $N$, we summarise performance by
averaging over seeds at fixed $N$ and then uniformly over the relevant range
of lengths. Over $N=10,\ldots,50$, FF-PCE obtains an average normalised merit
factor of $0.588$ before local search and $0.594$ after local search. IQP-PCE
obtains $0.283$ before local search and $0.485$ after local search. The small
postprocessing gain for FF-PCE indicates that its decoded sequences already
contain most of their final quality, whereas the larger improvement for
IQP-PCE shows that its final performance is more strongly affected by the
classical refinement step.

On the larger sizes $N=31,\ldots,50$, the postprocessed averages are $0.546$
for FF-PCE and $0.427$ for IQP-PCE. Applying the same local search to
uniformly random initial sequences gives an average $\rho_F=0.425$ on this
range. Thus, the large-size IQP-PCE postprocessed performance is comparable
to the random-start local-search baseline, whereas FF-PCE remains clearly
above it. This suggests that FF-PCE provides a useful correlation-based
initialisation for approximate LABS under the fixed, untuned training budget
considered here.

Overall, this experiment supports the same qualitative conclusion as the main
benchmarks: efficiently simulable PCE can generate useful optimisation
correlations beyond the graph and constraint problems studied in the main
text. At the same time, the LABS results should be interpreted only as an
approximate-solution-quality test, not as evidence for improved scaling in
finding exact LABS optima. Future experiments may explore the applications of efficiently-simulable PCE as exact solvers, to evaluate promising TTS  for LABS or similar problems.

\section{Considerations about the IQP ansatz}

\subsection{Number of gates to reach maximum expressivity}
\label{app:iqp_maximum_expressivity}

In this section, we characterise the maximum expressivity of the unitaries generated by the IQP ansatz.
By definition, these are contained in the space of all diagonal unitary matrices in the $X$ basis. Since conjugation by $H^{\otimes n}$ maps $X$ strings to $Z$ strings, we work in the computational basis throughout this section.
We show that every $n$-qubit unitary that is diagonal in the computational basis can, up to an overall global phase, be written as
\begin{equation*}
\prod_{\varnothing\neq T\subseteq[n]} e^{i\theta_T Z_T}.
\end{equation*}
First, note that the set $\{Z_T:T\subseteq[n]\}$ forms a basis for the diagonal Hermitian $2^n\times 2^n$ matrices. Let $D=\operatorname{diag}\!\left(e^{i\phi_x}\right)_{x\in\{0,1\}^n}$ be an arbitrary unitary diagonal in the computational basis. Choose one real representative $\phi_x$ for each diagonal phase and define $K=\operatorname{diag}\!\left(\phi_x\right)_{x\in\{0,1\}^n}$. Then $K$ is diagonal and Hermitian, so
\begin{equation*}
    K=\sum_{T\subseteq[n]}\alpha_T Z_T
\end{equation*}
for real coefficients $\alpha_T$. Since all the $Z_T$ commute,
\begin{equation*}
    D=e^{iK}\propto
    \prod_{\varnothing\neq T\subseteq[n]}e^{i\alpha_T Z_T}.
\end{equation*}
Conversely, every unitary in the product is diagonal, so this construction gives exactly the full family of diagonal unitaries, up to global phase. This family has real dimension $2^n-1$; hence an ansatz that parametrises all of it requires at least $2^n-1$ independent continuous parameters. The rotations generated by the $2^n-1$ nonidentity Pauli strings attain this bound. Conjugating the construction by $H^{\otimes n}$ gives the corresponding statement for IQP unitaries diagonal in the $X$ basis.

We conclude that the number of Pauli Z strings needed to reach maximum expressivity in the IQP ansatz is exponential. It is thus crucial to have a method to select a relevant subset of these strings to be included in the ansatz, as we do in \ref{subsec:iqp-pce}.

\subsection{Maximum number of independent variable assignments}
\label{app:iqp_maximum_independent_assignments}

When choosing the encoding for the problem variables into Pauli strings for IQP-PCE, one must ensure that the corresponding expectations are not overly constrained (or that such constraints respect those of the problem). For $n$ qubits, the maximum number of variables that can be assigned without algebraic constraints is $n$. For $k$ variables indexed by $1,\ldots,k$, define $m_j=\langle Z_{T_j}\rangle$. We call the assignment unconstrained if, using the maximally expressive IQP family, the vector $(m_1,\ldots,m_k)$ can be any element of $[-1,1]^k$. We identify each support $T_j$ with its incidence vector $t_j\in\mathbb{Z}_2^n$,
\begin{equation}
(t_j)_i = \begin{cases}
    1, \quad \text{if} \; i \in T_j, \\
    0, \quad \text{otherwise}
\end{cases}
\end{equation}
and write $Z_{t_j}$ for the corresponding Pauli string.

Linear independence of the $t_j$ is necessary. Indeed, if the vectors are dependent, there is a nonempty set $A\subseteq[k]$ such that
\begin{equation}
    \sum_{j\in A}t_j=0 \pmod 2,
    \qquad\text{and hence}\qquad
    \prod_{j\in A}Z_{t_j}=I.
\end{equation}
Every joint eigenvalue assignment $(s_1,\ldots,s_k)\in\{\pm1\}^k$ that can occur must therefore obey $\prod_{j\in A}s_j=1$. In particular, a vertex of the hypercube for which this product is $-1$ cannot be realised as an expectation vector. Thus the full cube $[-1,1]^k$ is not attainable. Linear independence also immediately implies $k\leq n$.

It remains to show that linear independence is sufficient, and that the required expectations can be realised by an IQP circuit. Since the $t_j$ are independent, there exist dual vectors $r_1,\ldots,r_k\in\mathbb{Z}_2^n$ satisfying
\begin{equation}
    t_j\cdot r_\ell=\delta_{j\ell}\pmod 2.
\end{equation}
For example, the $r_\ell$ can be chosen as the columns of a right inverse of the matrix whose rows are the $t_j$. Consider the IQP circuit
\begin{equation}
    U(\bm{\theta})=
    \prod_{\ell=1}^k e^{-i\theta_\ell X_{r_\ell}},
\end{equation}
where $X_{r_\ell}$ is the $X$ string with support $r_\ell$. To expand its action on the initial state, first note that
\begin{equation}
    e^{-i\theta_\ell X_{r_\ell}}
    =\cos\theta_\ell\,I-i\sin\theta_\ell\,X_{r_\ell},
\end{equation}
because $X_{r_\ell}^2=I$. For a binary vector $x\in\mathbb{Z}_2^n$, let $\ket{x}$ denote the corresponding computational-basis state. The action of an $X$ string is then
\begin{equation}
    X_{r_\ell}\ket{x}=\ket{x+r_\ell},
\end{equation}
where the addition is componentwise modulo two. In expanding the product defining $U(\bm{\theta})$, introduce $a_\ell\in\{0,1\}$ to record whether the identity term ($a_\ell=0$) or the $X_{r_\ell}$ term ($a_\ell=1$) is selected from the $\ell$-th factor. Since $\ket{0}^{\otimes n}=\ket{0}$ and the $X$ strings commute, the choice $a=(a_1,\ldots,a_k)$ produces the basis state $\ket{\sum_\ell a_\ell r_\ell}$ with amplitude $\prod_\ell(\cos\theta_\ell)^{1-a_\ell}(-i\sin\theta_\ell)^{a_\ell}$. Therefore,
\begin{equation}
    U(\bm{\theta})\ket{0}^{\otimes n}
    =\sum_{a\in\{0,1\}^k}
    \left[\prod_{\ell=1}^k
    (\cos\theta_\ell)^{1-a_\ell}(-i\sin\theta_\ell)^{a_\ell}\right]
    \ket{\textstyle\sum_\ell a_\ell r_\ell}.
\end{equation}
All sums in the ket labels are over $\mathbb{Z}_2^n$. They label distinct computational-basis states because the $r_\ell$ are linearly independent: if $\sum_\ell a_\ell r_\ell=\sum_\ell b_\ell r_\ell$, taking the inner product with $t_j$ gives $a_j=b_j$ for every $j$. Moreover, $Z_{t_j}$ has eigenvalue $(-1)^{a_j}$ on the basis state labelled by $a$, since
\begin{equation}
    t_j\cdot\left(\sum_{\ell=1}^k a_\ell r_\ell\right)
    =\sum_{\ell=1}^k a_\ell\delta_{j\ell}
    =a_j\pmod 2.
\end{equation}
It follows that
\begin{equation}
    \langle Z_{t_j}\rangle_{\bm{\theta}}
    =\cos^2\theta_j-\sin^2\theta_j
    =\cos(2\theta_j).
\end{equation}
Given any target $v\in[-1,1]^k$, choosing $\theta_j=\tfrac12\arccos(v_j)$ therefore yields $\langle Z_{t_j}\rangle=v_j$ for every $j$.

Consequently, linear independence over $\mathbb{Z}_2$ is necessary and sufficient for the observable assignment itself not to constrain the expectation vector under the maximally expressive IQP family. A restricted, problem-informed ansatz may impose additional constraints if it does not contain the required generators $X_{r_\ell}$. The simplest independent assignment is to map each variable to $Z_j$, with no more variables than available qubits, which is the choice made throughout the manuscript.

\end{document}